\documentclass[%
preprint 
amsmath,amssymb,
aps,
pra,
]{revtex4-2}

\usepackage{graphicx}

\usepackage{dcolumn}
\usepackage{bm}
\usepackage{longtable}
\usepackage{ulem}

\usepackage{xcolor,soul}
\usepackage{mathptmx}
\usepackage{color,amsmath,amssymb,amsfonts}%
\usepackage{ccaption}

\newcommand*\Laplace{\mathop{}\!\mathbin\bigtriangleup}

\begin{document}


\title{Turing patterns on non-fluctuating surfaces under mechanical stresses}

\author{Fumitake Kato}%
\email{katobnbf@ibaraki-ct.ac.jp}
\affiliation{
	National Institute of Technology (KOSEN), Ibaraki College, Hitachinaka, Japan
}
\author{Hiroshi Koibuchi} 
\email[Corresponding author: ]{koibuchih@gmail.com}
\affiliation{
	National Institute of Technology (KOSEN), Ibaraki College, Hitachinaka, Japan,\\
	and Argopilot Coorporation, Tomiya, Japan
}
\author{Madoka Nakayama }
\email{nakayama.madoka@tmd.ac.jp}
\affiliation{%
	Institute for Liberal Arts, Institute of Science Tokyo, Ichikawa, Japan
}%
\author{Sohei Tasaki}
\email{tasaki@math.sci.hokudai.ac.jp}
\affiliation{%
	Department of Mathematics, Faculty of Science, Hokkaido University, Sapporo, Japan
}%
\author{Tetsuya Uchimoto}
\email{uchimoto@tohoku.ac.jp}
\affiliation{%
	Institute of Fluid Science (IFS), Tohoku University, Sendai, Japan,
}%
\affiliation{ ELyTMaX, CNRS-Universite de Lyon-Tohoku University, Sendai, Japan.}

\begin{abstract}
This paper presents a numerical study of Turing patterns (TPs) governed by reaction–diffusion equations for the activator $u$ and the inhibitor $v$ on two- and three-dimensional lattices without vertex fluctuations. In this framework, $u$ and $v$ are fixed at discrete spatial locations, as pigment cells on zebrafish skin or  shell patterns. Mechanical effects are incorporated through the Finsler geometry modeling formulation, which introduces an internal degree of freedom, $\vec{\tau}$, representing the direction of mechanical stress. A tensile-stress formula based on the Gaussian bond potential is shown to be well defined on non-fluctuating lattices, enabling the entropy associated with stress relaxation to be evaluated in a manner analogous to that on fluctuating surfaces. The results indicate that biological TPs respond to external mechanical forces in much the same way as TPs on fluctuating membranes. Simulation codes are provided in the Supplementary Material.
\end{abstract}

\maketitle
\section{Introduction\label{intro}}
\begin{figure}[h!]
	\centering{}\includegraphics[width=10.5cm]{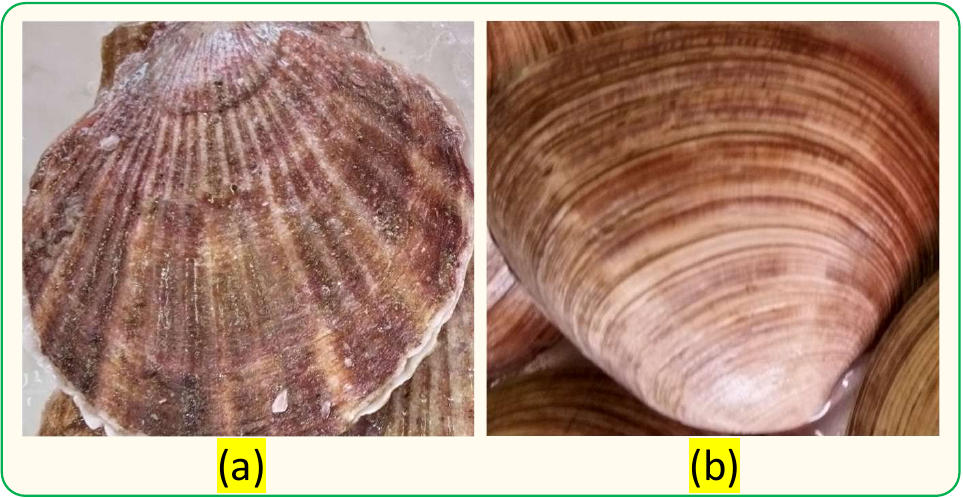}
	\caption{  Shell patterns that emerge during development can also be understood as an example of TP on solid materials, similar to TPs on soft materials, such as zebras and fish.   \label{fig-1}  }
\end{figure}

A substantial body of research has been devoted to mathematical models of biological pattern formation in systems such as zebras, fish, and seashells \cite{Meinhardt-ACPress1982,Cross-Hohenberg-RMP1993,Koch-Meinhardt-RMP1994,Meinhardt-AlgoBeauty} (Fig. \ref{fig-1}). These patterns are widely understood as manifestations of reaction–diffusion dynamics involving local enhancement and long-range inhibition \cite{Taylor-Tinsley-NatChem2009,Falasco-etal-PRL2018,Avanzini-etal-JCP2024}. In standard RD models, the activator and inhibitor concentrations are represented by two scalar fields, $u$ and $v$, respectively \cite{FitzHugh-BP1961,Nagumo-etal-ProcIRE1962,Gierer-Meinhardt-Kyber1972}. Anisotropic Turing patterns generally require anisotropic diffusion coefficients in addition to a sufficiently large inhibitor diffusivity relative to that of the activator \cite{Kondo-Nature1995,Kondo-Miura-Science2010,Shoji-etal-DevDyn2003,Iwamoto-Shoji-RIMS2018,Sekimura-etal-PRSL2000,Shoji-Iwasa-Forma2003}. Such patterns have been reported not only in biological systems \cite{Nakamasu-etal-PNAS2009,Mahalwar-etal-Science2014,Yamanaka-etal-PNAS2014,Sawada-etal-GCell2018} but also in non-living materials \cite{Tan-etal-Science2018,Fuseya-etal-NatPhys2021,Noble-etal-PRL2020} and complex networks \cite{Othmer-Scriven-JTB1971,Nakao-etal-NatPhys2010,Asllani-etal-Ncom2014,Carletti-Nakao-PRE2020,Asllani-etal-NatCom2014,Asllani-etal-PRE2014,Petit-etal-PhysA2016}. Although the present study focuses on biological Turing patterns, the proposed framework is sufficiently general to be applied to non-living systems as well.

In order to ascertain the origin of the anisotropic diffusion ($D_{u,v}^x\!\not=\!D_{u,v}^y$), the ordinary mathematical model described by the RD equation is extended to a hybrid simulation model by including internal degree of freedom (IDOF) ($\vec {\tau} \in S^2/2$: half unit sphere) \cite{Diguet-etal-PRE2024,FKato-SM2025}, in the framework of Finsler geometry (FG) modeling technique \cite{Koibuchi-Sekino-PhysA2014,Proutorov-etal-JPC2018,Diguet-etal-CMS2024,SS-Chern-AMS1996}. 
The $\vec {\tau}$ denotes the "microscopic stress direction" representing direction of cell movement due  to external stimuli, such as mechanical forces. 	
{ This IDOF $\vec{\tau}$ represents a lattice deformation or strain direction at each lattice vertex on the fixed lattice model. In this sense, the mean value of $\vec{\tau}$ corresponds to the diagonal part of the stress tensor. Additionally, the mean value of uniformly aligned microscopic $\vec{\tau}$ determines length scales for the interaction between two neighboring $u, v$ on the surfaces, and the  $\vec {\tau}$ direction signifies the orientation of polymeric substances or the movement of chemical substances on membranes, whether polymerized or fluid \cite{KANTOR-NELSON-PRA1987,Gompper-Kroll-PRA1992,HELFRICH-1973,Polyakov-NPB1986,Peliti-Leibler-PRL1985,Bowick-PRep2001,NELSON-SMMS2004,Wheater-JPA1994,KOIB-PRE-2005}. Therefore, given that the concentration fields $u$ and $v$ remain constant in the absence of movement, it is reasonable to hypothesize the mobility of chemical substances in the biological TP system \cite{Peleg-etal-PlosOne2011,Wu-etal-NatCom2018,Tamemoto-Noguchi-SciRep2020,Tamemoto-Noguchi-SM2021,Tamemoto-Noguchi-PRE2022,Noguchi-SciRep2023} with curvature effects \cite{Orlandini-SM2013,Varea-etal-PRE1999,Krause-etal-BulMatBiol2019,Krause-etal-PhiLTrRSoc2021,Nishide-Ishihara-PRL2022}.

However, Bullara et al. have suggested that the cell movement may not be indispensable for the formation of the Turing pattern in zebrafish \cite{Bullara-Decker-NatCom2015}. A notable distinction emerges when comparing their mathematical modeling to the RD equation of Turing, despite the shared conceptual foundation of "local enhancement and long-range inhibition." Indeed, the chemical substances, typically described by $u$ and $v$, were identified as pigments in zebrafish pattern \cite{Nakamasu-etal-PNAS2009,Mahalwar-etal-Science2014,Yamanaka-etal-PNAS2014,Sawada-etal-GCell2018}.  Pigment cells are regarded as discrete entities, created, exchanged and annihilated at a fixed position, despite some movement being observed \cite{Yamanaka-etal-PNAS2014,Sawada-etal-GCell2018}. 
Therefore,  it is of considerable interest to investigate whether TPs can be identified as well-defined objects on non-fluctuating (fixed) lattices using the same methodology, including entropy calculations, that has been applied to fluctuating membranes in Refs. \cite{Diguet-etal-PRE2024,FKato-SM2025}. Although entropy can be readily evaluated for fluctuating lattices, its evaluation on non-fluctuating lattices remains challenging. Since entropy is closely associated with the relaxation of IDOF, which governs the orientation of TPs, the inability to evaluate entropy on non-fluctuating lattices would preclude establishing the relationship between TPs and IDOF relaxation.

The aim of this study is to determine whether both the relationship between TPs and IDOF relaxation and the proposed origin of TPs, namely mechanical surface deformation, remain unchanged when moving from fluctuating to non-fluctuating lattices.
In this paper, we extend the hybrid simulation model of TPs for membranes on fluctuating lattices to a fixed lattice model by discarding the vertex fluctuation. The fixed lattice model is defined on two-dimensional (2D) square and  triangular lattices, and three-dimensional (3D) cubic lattices, all of which are free of vertex fluctuation. The lattice structure is regarded as an interaction network of pigment cells.  The concept of biological patterns is a two-dimensional object, and it does not involve any 3D structure. Therefore, the utilization of 3D models is not an indispensable requirement for TP modeling. To this end, a thin plate is employed for the 3D model, and the implemented mechanical property is examined to check its universality in the sense of dimension independence.  If TPs exhibit different behavior in 3D and 2D models, this would imply that dynamical anisotropy is sensitive to dimensionality, even though a thin 3D plate is physically equivalent to a 2D surface.  The variables assumed in the  model are $u,v$ and $\vec{\tau}$. The vertex position $\vec{r}$ of lattices is fixed and not included as a variable, although it is used  to define the Finsler metric. The inclusion of the IDOF $\vec{\tau}$  is predicated on the expression of a mechanical elastic property in materials such as zebrafish and sea shells. The variables $u,v$ can be extended to three different ones for pigment cells as in Refs.  \cite{Nakamasu-etal-PNAS2009,Mahalwar-etal-Science2014,Sawada-etal-GCell2018}.  However,  we employ the same $u,v$ as those in Refs. \cite{Diguet-etal-PRE2024,FKato-SM2025} to solely elucidate the influence of fixing the cell position $\vec{r}$.

It  must be emphasized that the fixed lattice model under consideration is physically nontrivial, despite the fact that the  model is more easily simulated than the fluctuating lattice models in Refs. \cite{Diguet-etal-PRE2024,FKato-SM2025}. The underlying reason for the non-triviality is that entropy can only be calculated on fluctuating lattices by using the scale invariant property of the partition function \cite{Wheater-JPA1994}. In contrast, this property is not present in the fixed lattice model, and therefore, the fixed lattice model should be defined in the limit of small fluctuations of the fluctuating models. In the definition procedure of the fixed lattice model,  the elastic energy of the lattice network is included in the model Hamiltonian, as in the fluctuating membranes, despite the constant lattice bond length.  The implemented elastic energy allows us to obtain the small fluctuation limit of a mechanical free energy. Using this free energy and microscopic Hamiltonian, we can calculate entropy like in the fluctuating membrane models in  Ref. \cite{FKato-SM2025}. The supplementary material provides a detailed discussion of the non-triviality of the fixed-lattice modeling.

This paper is organized as follows: The main text introduces the aforementioned extended model and  presents the results obtained from fixed 2D regular square and regular triangular lattices, as well as from a fixed 3D cubic lattice. The primary objective of this study is to examine how TPs respond to external mechanical stimuli,  particularly lattice deformations, in order to ascertain whether the responses are the same as those of fluctuating lattices. Additionally, it addresses whether the stress relaxation phenomena can be adequately captured in the canonical simulations. To this end, the stress formula, which is typically derived from fluctuating lattices \cite{Diguet-etal-PRE2024,FKato-SM2025}, is applied to fixed lattices.  The validity of the stress formula on the fixed square and triangular lattices is demonstrated by presenting the details of this problem in the supplementary material. 
The supplementary material provides simulation codes for generating converged configurations on non-fluctuating square, triangular, and 3D cubic lattices.

	\begin{figure}[h!]
	\centering{}\includegraphics[width=10.5cm]{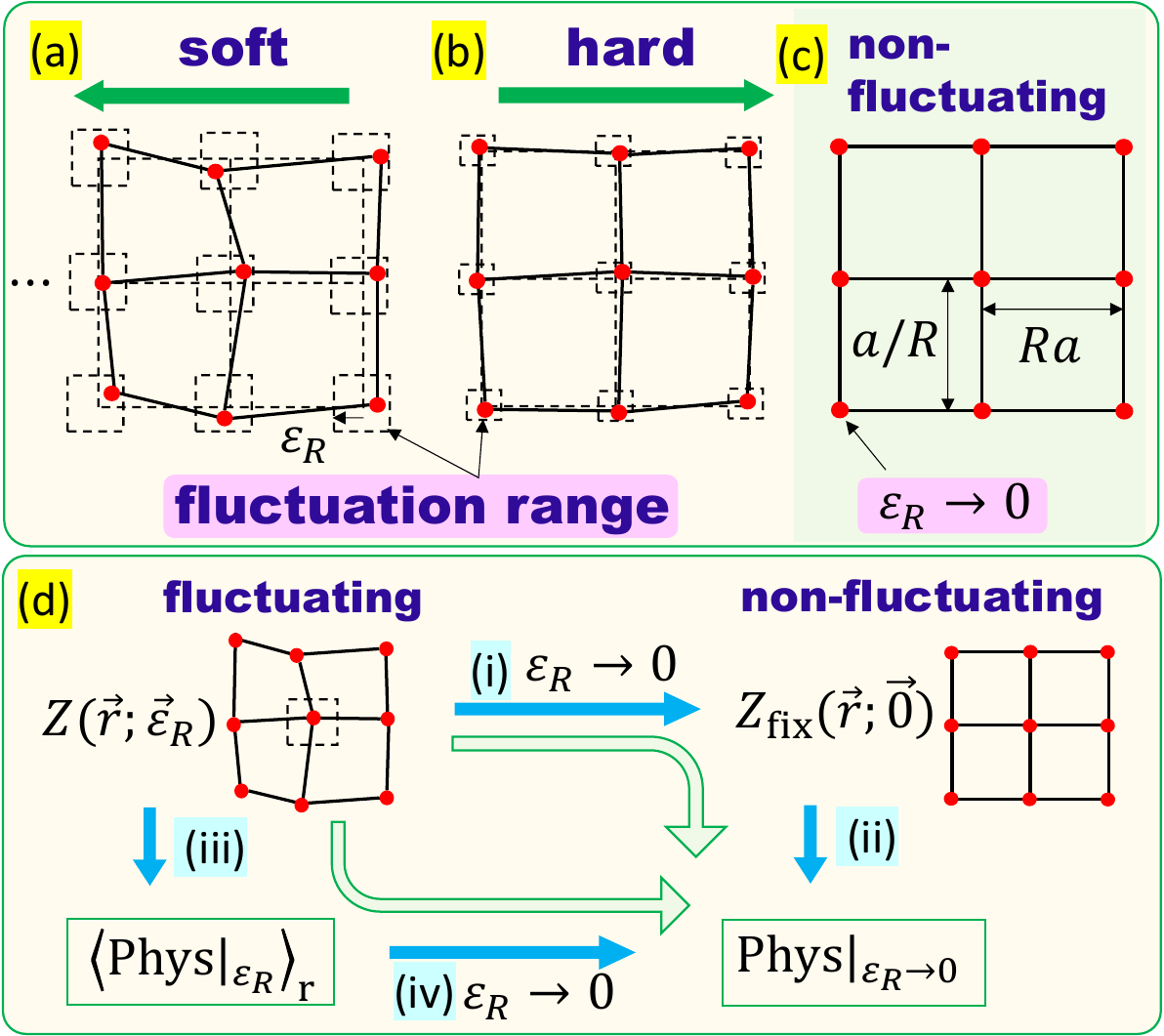}
	\caption{The range  $\varepsilon_R$ of the vertex fluctuations defines the material property:  (a) soft ($\Leftrightarrow$ large $\varepsilon_R$) and (b) hard ($\Leftrightarrow$ small $\varepsilon_R$), and (c) fixed lattice  ($\Leftrightarrow$ $\varepsilon_R\!\to\!0$) with a deformed lattice spacing $(Ra,a/R)$ with a deformation parameter $R$ and a constant $a$.  (d) An illustration of the compatibility relation for  the zero fluctuation limit  $\varepsilon_R\!\to\!0$ and the calculation of physical quantity, where $Z$ and $Z_{\rm fix}$  are the partition functions of the fluctuating and fixed lattices, respectively. \label{fig-2}  }
\end{figure}

\section{Non-fluctuating Square, triangular and cubic lattices \label{2D-lattice}}
In this section, we present a detailed information of 2D and 3D lattices, on which the discrete RD equation and the model Hamiltonian, introduced in the following section, are defined.

Before presenting the lattice construction, we explain why the model on the non-fluctuating  lattice is physically non-trivial. If the non-fluctuating lattice model is regarded as the zero-fluctuation limit of the fluctuating lattice model, entropy remains well defined and can, in principle, be evaluated for the non-fluctuating system. In this case, the relaxation of the IDOF, $\vec{\tau}$, also retains a clear physical meaning. It is therefore important to clarify the connection between fluctuating and non-fluctuating lattice models. Figures \ref{fig-2}(a)--(c) illustrate fluctuating lattices with both  large and small fluctuation ranges $\varepsilon_R$, as well as a non-fluctuating lattice with a deformed lattice spacing $(Ra,a/R)$, where $R$ and $a$ are a deformation parameter and a constant lattice spacing, respectively.  In the context of fluctuating lattices, it is hypothesized that the range of fluctuations is represented by a small square centered at the fixed vertex position of the non-fluctuating lattice. 
In the model with vertex fluctuations, the Gaussian bond potential (GBP), defined by the sum of the bond length squares, is assumed as in the linear chain model \cite{MDoi-Edwards-1986,MDoi-2013}. Consequently,  the model describes soft materials for a finite non-zero value of $\varepsilon_R$. Conversely,  the model describes hard materials in the limit of zero fluctuation $\varepsilon_R\!\to\!0$ due to the frozen degrees of freedom for the vertex position.

To explain this issue in detail, we employ the concept of the partition function in surface modeling. The partition function $Z$ for the Monte Carlo (MC) update of $\vec{\tau}$ is given by
\begin{eqnarray}
	\label{part-funct}
	Z_{\rm fix}=\int\prod_{i=1}^N d\vec{\tau}_i\exp\left(-H(\vec{r},\vec{\tau})/T\right)=\sum_{\vec{\tau}} \exp\left(-H(\vec{r},\vec{\tau})/T\right),
\end{eqnarray}
where $\vec{\tau}\in S^1/2$ for the 2D models and $\vec{\tau}\in S^2/2$ for the 3D model. Here, $T$ denotes the temperature and is fixed at 1 in all simulations. The Boltzmann constant $k_B$ is also set to unity. The total Hamiltonian $H(\vec{r},\vec{\tau})$ is introduced below.

The relation between the partition functions on the fluctuating and non-fluctuating lattices is illustrated at the process (i) in Fig. \ref{fig-2}(d). The partition function $Z_{\rm fix}$ on the non-fluctuating lattice is obtained from $Z$ on the fluctuating lattice in the limit of $\varepsilon_R\!\to\!0$. Let $\left.{\rm Phys}\right|_{\varepsilon_R=0}$ and $\langle\left.{\rm Phys}\right|_{\varepsilon_R} \rangle_{\rm{r}}$  be the formula for physical quantity obtained by using $Z_{\rm fix}$ and $Z$, as illustrated at the processes (ii) and (iii) in  Fig. \ref{fig-2}(d), respectively. The symbol $\langle \cdots \rangle_{\rm{r}}$ in $\langle\left.{\rm Phys}\right|_{\varepsilon_R} \rangle_{\rm{r}}$ denotes the mean value operation under the lattice fluctuation.  The mean value operation $\langle\cdots\rangle_\tau$ for the IDOF, which is different from $\langle \cdots \rangle_{\rm{r}}$, is not included in $\left.{\rm Phys}\right|_{\varepsilon_R=0}$ or $\langle\left.{\rm Phys}\right|_{\varepsilon_R} \rangle_{\rm{r}}$, for simplicity. 
The non-trivial problem is whether the following relation is satisfied:
\begin{eqnarray}
	\label{nontrivila-phys-relation}
\lim_{\varepsilon_R\to 0}\langle\left.{\rm Phys}\right|_{\varepsilon_R} \rangle_{\rm{r}}= \left.{\rm Phys}\right|_{\varepsilon_R= 0}.
\end{eqnarray}
One example of  $\left.{\rm Phys}\right|_{\varepsilon_R=0}$ is the surface tension or tensile stress, which is indispensable for calculating free energy and entropy, of which the simulation data will be presented in an upcoming section on presentations. It is important to note that the partition function $Z_{\rm fix}$ does not include the integration of vertex positions. However, the tensile stress formula is known to be only well defined in the presence of the positional integration on fluctuating lattices in the FG modeling framework on the basis of GBP. Therefore, ascertaining whether the stress can be calculated on the non-fluctuating lattice without the positional integration is a non-trivial aspect of our modeling. 
 It should be noted that the relation of Eq. (\ref{nontrivila-phys-relation}) is not always necessary for physical quantities other than the tensile stress.  A comprehensive examination of this issue described in Eq. (\ref{nontrivila-phys-relation})  can be found in the supplementary material.

\begin{figure}[h!]
\centering{}\includegraphics[width=10.5cm]{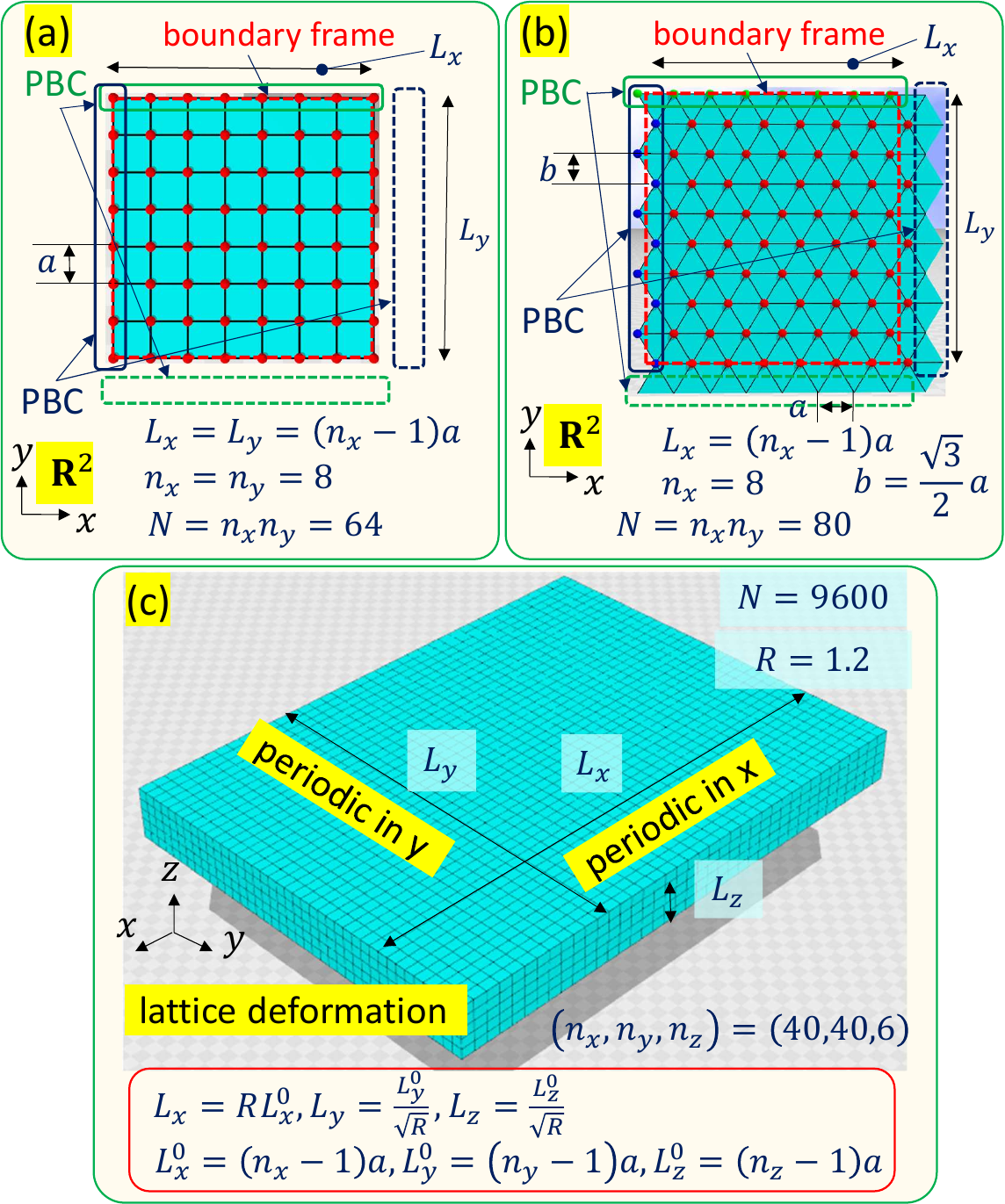}
\caption {  The regular square and triangular lattices (a) and (b), respectively,  where the total number of vertices is small, are plotted to elucidate the underlying structure. (c) The 3D cubic lattice of side lengths $(L_x,L_y,L_z)$, where the total number of vertices is $N\!=\!9600$, is utilized in the simulations. The 2D lattices employ periodic boundary conditions (PBCs), while the 3D lattice's PBCs are applied exclusively in the $x$, $y$ directions. 
\label{fig-3} }
\end{figure}
Now, return to the lattice construction. A two-dimensional triangulated lattice of size  $N\!=\!n_xn_y\!=\!80$, which is the total number of vertices, and a square lattice of size $N\!=\!n_xn_y\!=\!64$  are illustrated in Figs.\ref{fig-3}(a) and \ref{fig-3}(b). The 3D cubic lattice of size $N\!=\!n_xn_yn_z\!=\!9600$ is plotted in Fig. \ref{fig-3}(c), which will be used in the simulations. 

The plate is deformed along the $x$ axis in the simulations, and the side lengths  $(L_x,L_y)$ of 2D models are given by
\begin{eqnarray}
\label{side-length-2D}
L_x=R L_x^0, \quad L_y=\frac{L_y^0}{R},\quad (L_x^0,L_y^0)=((n_x-1)a,(n_y-1)a), \;
		a=\left\{ \begin{array}{@{\,}ll}
	0.58 \;({\rm triangle})\\
	0.72 \; ({\rm square})
\end{array}\right.,
\end{eqnarray}
with the deformation ratio $R$,  and $(L_x,L_y,L_z)$ of 3D model are
\begin{eqnarray}
	\label{side-length-3D}
\begin{split}
 &L_x=R L_x^0, \;L_y=\frac{L_y^0}{\sqrt{R}},\;L_z=\frac{L_z^0}{\sqrt{R}},\\ &(L_x^0,L_y^0,L_z^0)=((n_x-1)a,(n_y-1)a,(n_z-1)a),\quad a=0.73 \; ({\rm cube}),
\end{split}
\end{eqnarray}
where $L_\mu^0, (\mu=x,y,z)$ are the side lengths of nondeformed lattice corresponding to $R\!=\!1$, and $a$ is the lattice spacing which defines the simulation unit of length. The values $a$ in Eqs. (\ref{side-length-2D}) and (\ref{side-length-3D}) are discussed below.

The ratio $R>1$ ($R<1$) describes an extension (compression) along the $x$ axis or, equivalently, a compression (extension) along the $y$ direction. In this case, the area or volume of the plate  remains unchanged despite the extension (compression): $L_xL_y\!=\!L_x^0L_y^0$ for the 2D lattices and  $L_xL_yL_z\!=\!L_x^0L_y^0L_z^0$ for the 3D lattice.

\section{Finsler geometry modeling of Turing patterns\label{FG-modeling}}
\subsection{FitzHugh-Nagumo equation}
 The TPs are described by the equations for the variables $u(x,y)$ and $v(x,y)$ on the 2D plates such that
\begin{eqnarray}
	\begin{split}
		\label{FN-eq}
		&\frac {\partial u}{\partial t}=D_u {\Laplace}(\tau)u + f(u ,v), \quad  f=u -u ^3-v, \\
		&\frac {\partial v}{\partial t}=D_v  \Laplace(\tau) v + g(u,v), \quad g= \gamma(u -\alpha v).
	\end{split}
\end{eqnarray}
These are known as the FitzHugh-Nagumo equations \cite{Kondo-Nature1995,Shoji-etal-DevDyn2003,Iwamoto-Shoji-RIMS2018,Sekimura-etal-PRSL2000,Shoji-Iwasa-Forma2003,Kondo-Miura-Science2010}. The diffusion coefficients $D_u$ and $D_v$ and  the parameters $\gamma$ and $\alpha$  are  appropriately fixed in the simulations.  In the standard Euclidean expression, the Laplace operator ${\Laplace}$ is given by  ${\Laplace}\!=\!\frac{\partial^2}{\partial x^2}\!+\!\frac{\partial^2}{\partial y^2}$ for 2D case. The operator ${\Laplace}(\tau)$ in this paper depends on the IDOF $\vec{\tau}$, and the discrete expression of ${\Laplace}(\tau)$ is provided in Appendix \ref{App-A}.

\subsection{Hamiltonian for Turing patterns \label{model-Hamiltonian}}
The combined system of TPs and the material is defined by the variables $(u,v)$ and IDOF $\vec{\tau} (\in S^1/2: {\rm half\; circle})$ for 2D case and  $\vec{\tau} (\in S^2/2: {\rm half\; sphere})$ for 3D case  \cite{note-1}. As mentioned in the introduction, the IDOF $\vec{\tau}$ corresponds to the diagonal part of the stress tensor and  represents the direction of stress. The non-diagonal part, or shear stress, is not used. Note that the macroscopic stress tensor is not calculated directly. Instead, the diagonal part is incorporated in part as a microscopic variable IDOF  in the modeling. 
The vertex position $\vec{r}_i (\in {\bf R}^D)$, $(D\!=\!2,3)$ is fixed, while the $\vec{\tau}_i$ is updated by a Metropolis Monte Carlo (MC) technique \cite{Metropolis-JCP-1953,Landau-PRB1976}. The discrete Hamiltonian used in MC for the update of $\vec{\tau}_i$ is given by
\begin{eqnarray}
	\label{discrete-Hamiltonian}
	\begin{split}
	&H(\vec{r},\vec{\tau})=	 H_1+\lambda H_\tau +\left(H_u-\frac{1}{\gamma}H_{v}\right),\\
	&H_1=\sum_{ij}\Gamma_{ij}^G(\tau)\ell_{ij}^2,\quad \ell_{ij}=\|\vec{r}_j-\vec{r}_i\|,\quad
	H_\tau=\frac{3}{2}\sum_{ij}\left(1-(\vec{\tau}_i\cdot\vec{\tau}_j)^2\right),\\  
&\Gamma_{ij}^G=\left\{ \begin{array}{@{\,}ll}
	{\rm sine\; type}\quad  ({\rm model\; 1}) \\
	{\rm cosine\; type}\quad  ({\rm model\; 2})
\end{array} \right..
	\end{split}
\end{eqnarray}
The terms in $\left(H_u\!-\!\frac{1}{\gamma}H_{v} \right)$  are relatively small compared to the other terms, however, they are not negligible. 
The RD equations in Eq. (\ref{FN-eq}) are obtained from a variational treatment of $H_u\!-\!\frac{1}{\gamma}H_v$ (Appendix \ref{App-B}). If the contribution of this term is large compared with other contributions, such as $H_1\!+\!\lambda H_\tau$, then the anisotropic diffusion of $u$ and $v$,  characterized by the coefficients $D_{ij}^{u,v}(\tau)$ of $H_{u,v}^D$, may affect the orientation of $\vec{\tau}$. However, since the contribution of  $\left(H_u\!-\!\frac{1}{\gamma}H_{v} \right)$ is small, the orientation of $\vec{\tau}$ is governed primarily by $H_1\!+\!\lambda H_\tau$. The resulting orientation then gives rise to anisotropic diffusion of $u$ and $v$ through the diffusion operators $\Laplace(\tau)u$ and  ${\Laplace}(\tau)v$  appearing in the RD equations (\ref{FN-eq}). The discrete representation of  ${\Laplace}(\tau)u$ is given in Eq. (\ref{discrete-Laplace-App}).

The first term $H_1$, which is the GBP, is defined by the sum of bond length squares $\ell_{ij}^2$, which are fixed, while the intensive part $\Gamma_{ij}^G(\tau)$ varies depending on $\vec{\tau}$. 
We briefly outline the discretization procedure for $H_1$ here. Additional details are given in Appendices \ref{App-A} and \ref{App-C}, where some of the expressions presented below are reproduced for completeness. The continuous form of $H_1$ is given by
\begin{eqnarray}\label{continuous-H1-main}
	H_1=\int \sqrt{g^G}d^2x g_G^{ab}\frac{\partial \vec{r}}{\partial x^a}\cdot\frac{\partial \vec{r}}{\partial x^b}, 
\end{eqnarray}
 where $\vec{r}(\in {\bf R}^2)$ is a position vector, and $2\times 2$ matrix $g_G^{ab}$ is the inverse of the Finsler metric $g_{ab}^G$.   The discrete  form of $g^G_{ab}$ with the local coordinate  system $(x,y)$ at vertex $i$ (Fig. \ref{fig-4}(a)) is given by
\begin{eqnarray}
	\label{F-metric-main}
	\begin{split}
		&g_{ab}^G=\begin{pmatrix}
			(\chi_{i1}^G)^{-2} & 0\\
			0 & (\chi_{i2}^G)^{-2} 
		\end{pmatrix},\\
		&\sqrt{g^G}=\sqrt{\det g_{ab}^G}=\frac{1
		}{\chi_{i1}^{G}\chi_{i2}^{G}}, \quad g^{ab}_G=(g_{ab}^G)^{-1}.
	\end{split}
\end{eqnarray}
 Because the components of $g_{ab}^G$ are defined only at lattice vertices, the metric $g_{ab}^G$  should be regarded as a discrete quantity. The unit Finsler length $\chi^G_{i1}$, which enters the definition of $g_{11}^G$, is defined by
\begin{eqnarray}
	\label{discrete-chi}
	\begin{split}
		\chi_{i1}^G=\left\{ \begin{array}{@{\,}ll}
			\sqrt{1-(\vec{\tau}_i\cdot\vec{e}_{i1})^2}+\chi_0\quad  (\sin)\; ({\rm for\; model\; 1}) \\
			|\vec{\tau}_i\cdot\vec{e}_{i1}|+\chi_0\qquad\qquad  (\cos)\; ({\rm for\; model\; 2}) 
		\end{array} \right.,\quad \chi_0=1.
	\end{split}
	\end{eqnarray}
	\begin{figure}[h!]
		\centering{}\includegraphics[width=8.5cm]{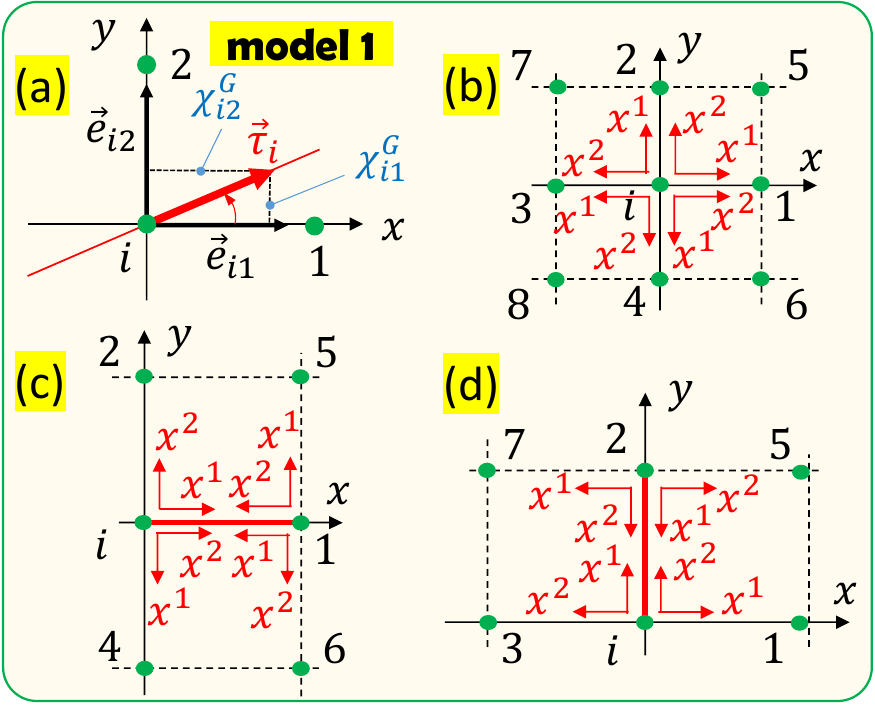}
		\caption{
(a) A local coordinate system $(x,y)$ at vertex $i$ and definitions of $\chi_{ij}^G, (j=1,2)$ using the IDOF $\vec{\tau}_i$ and the basis vectors $\vec{e}_{ij}, (j=1,2)$, where $\chi_0\!=\!0$ is assumed, (b) four possible local coordinate systems $(x^1,x^2)$ at vertex $i$, and four possible local coordinate systems along (c) bond $i1$ and (d) bond $i2$.   \label{fig-4}  }
	\end{figure}
Replacing the integral with the sum over vertices $\int \sqrt{g^G}d^3x\!\to\!\sum_i \sqrt{g^G}\!=\!\sum_i\frac{1
}{\chi_{i1}^{G}\chi_{i2}^{G}}$, and the differentials with the difference operations $g_G^{11}\frac{\partial \vec{r}_i}{\partial x}\cdot \frac{\partial \vec{r}_i}{\partial x}\!\to\!(\chi_{i1}^{G})^2\left(\vec{r}_1\!-\!\vec{r}_i\right)^2\!=\!(\chi_{i1}^{G})^2\ell_{i1}^2$, $g_G^{22}\frac{\partial \vec{r}_i}{\partial y}\cdot \frac{\partial \vec{r}_i}{\partial y}\!\to\!(\chi_{i2}^{G})^2\left(\vec{r}_2\!-\!\vec{r}_i\right)^2\!=\!(\chi_{i2}^{G})^2\ell_{i2}^2$, where $\ell_{ij}^2\!=\!\left(\vec{r}_i\!-\!\vec{r}_j\right)^2$ is utilized, we obtain the discrete form of $H_1$ such that $H_1\!=\!\sum_i  \left(\frac{\chi_{i1}^{G}
}{\chi_{i2}^{G}}\ell_{i1}^2+\frac{\chi_{i2}^{G}
}{\chi_{i1}^{G}}\ell_{i2}^2\right)$. 
As shown in Fig \ref{fig-4}(b), there are four possible local coordinates at $i$. Therefore, it is reasonable to include the other three possible contributions to $H_1$ at $i$ such that     
\begin{eqnarray}
	\label{discrete-H1-vertex-main}
	\begin{split}
		H_1&\to
	\sum_i  \left(
	\frac{\chi_{i1}^{G}}{\chi_{i2}^{G}}\ell_{i1}^2 
	+\frac{\chi_{i2}^{G}}{\chi_{i1}^{G}}\ell_{i2}^2 
	+\frac{\chi_{i2}^{G}}{\chi_{i3}^{G}}\ell_{i2}^2
	+\frac{\chi_{i3}^{G}}{\chi_{i2}^{G}}\ell_{i3}^2 
	+\frac{\chi_{i3}^{G}}{\chi_{i4}^{G}}\ell_{i3}^2
	+\frac{\chi_{i4}^{G}}{\chi_{i3}^{G}}\ell_{i4}^2 
	+\frac{\chi_{i4}^{G}}{\chi_{i1}^{G}}\ell_{i4}^2
	+\frac{\chi_{i1}^{G}}{\chi_{i4}^{G}}\ell_{i1}^2\right)\\
	&=\sum_i  \left(
\left[\frac{\chi_{i1}^{G}}{\chi_{i2}^{G}}+\frac{\chi_{i1}^{G}}{\chi_{i4}^{G}}\right]\ell_{i1}^2 
+\left[\frac{\chi_{i2}^{G}}{\chi_{i1}^{G}}+\frac{\chi_{i2}^{G}}{\chi_{i3}^{G}}\right]\ell_{i2}^2
+\left[\frac{\chi_{i3}^{G}}{\chi_{i2}^{G}}+\frac{\chi_{i3}^{G}}{\chi_{i4}^{G}}\right]\ell_{i3}^2 
+\left[\frac{\chi_{i4}^{G}}{\chi_{i3}^{G}}+\frac{\chi_{i4}^{G}}{\chi_{i1}^{G}}\right]\ell_{i4}^2 
\right)
	\end{split}
\end{eqnarray}
It is also convenient to replace the sum over vertex with the sum over bonds in the expression of $H_1$. For bonds $i1$ and $i2$, we have four possible local coordinate systems (Figs. \ref{fig-4}(c),(d)).  Thus, for the square lattice, $H_1$ can be written as 
\begin{eqnarray}
	\label{discrete-H1-bonds-main}
	\begin{split}
	&	H_1=\sum_{ij} \Gamma_{ij}^G\ell_{ij}^2, \\
	&		\Gamma_{i1}^G=\bar{\Gamma}_G^{-1}       \left(\frac{\chi_{i1}^{G}}{\chi_{i2}^{G}}\!+\!\frac{\chi_{i1}^{G}}{\chi_{i4}^{G}}\!+\!\frac{\chi_{1i}^{G}}{\chi_{15}^{G}}\!+\!\frac{\chi_{1i}^{G}}{\chi_{16}^{G}}\right), 	
		\Gamma_{i2}^G=\bar{\Gamma}_G^{-1} \left(\frac{\chi_{i2}^{G}}{\chi_{i1}^{G}}\!+\!\frac{\chi_{i2}^{G}}{\chi_{i3}^{G}}\!+\!\frac{\chi_{2i}^{G}}{\chi_{25}^{G}}\!+\!\frac{\chi_{2i}^{G}}{\chi_{27}^{G}}\right), 
\end{split}
\end{eqnarray}
where the symbol $\bar{\Gamma}_G$ is a normalization factor chosen such that $\Gamma_{ij}^G(\tau^{\rm iso})\to 1$ for isotropic $\tau^{\rm iso}$, regardless of the value of $\chi_0$ (Appendix \ref{App-C}). Note that $\chi_{i2}^G\!=\!\chi_{i4}^G$ and $\chi_{15}^G\!=\!\chi_{16}^G$ in $\Gamma_{i1}^G$, and  $\chi_{i1}^G\!=\!\chi_{i3}^G$ and $\chi_{25}^G\!=\!\chi_{27}^G$ in $\Gamma_{i2}^G$.

The extensive part  $\ell_{ij}^2$ in $H_1$ changes uniformly under the lattice deformation controlled by $R$. Through the intensive part  $\Gamma_{ij}^G(\tau)$, this change affects the orientation of $\vec{\tau}$. Because $\Gamma_{ij}^G(\tau)$ depends on model-specific definition of    $\chi_{ij}^G$ [Eq. (\ref{discrete-chi})], models 1 and 2 favor different orientation of  $\vec{\tau}$.  Consequently, the TP direction becomes either parallel or antiparallel to  $\vec{\tau}$.
 
 The value of $\chi_0$ determines the degree of mechanical anisotropy.  In the limit $\chi_0\!\to\! 0$, $\chi_{ij}^G$ may approach zero in both models 1 and 2. Nevertheless, configurations with vanishing $\chi_{ij}^G$ are excluded by the rational dependence of $\Gamma_{ij}^G$  in Eq. (\ref{discrete-H1-bonds-main}). In the opposite limit  $\chi_0\!\to\!\infty$, $\chi_{ij}^G\!\to\! 1$, and 
 the anisotropy disappears. We therefore adopt the intermediate value $\chi_0\!=\!1$, which is twice the value used in Refs. \cite{Diguet-etal-PRE2024,FKato-SM2025}. Remarkably, despite this quantitative difference, both sets of models exhibit nearly identical TP responses to lattice deformation.
 
  The second term $\lambda H_\tau$ with the coefficient $\lambda$ describes the nearest neighbor correlation of non-polar variable $\vec{\tau}$ ($\Leftrightarrow \vec{\tau}\in S^1/2$ or $\vec{\tau}\in S^2/2$) and is defined to be positive.

The Hamiltonians $H_u$ and $H_v$ are given by
\begin{eqnarray}
	\label{discrete-Hu-Hv}
	\begin{split}
		&H_u=D_uH_u^D+H_u^R,\quad H_u^D=\sum_{ij}D_{ij}^u(\tau)\left(u_i-u_j\right)^2,\quad H_u^R=-\sum_i \left(u_i^2-\frac{u_i^4}{2}-u_iv_i\right)\\ 
		&H_v=D_vH_v^D+H_v^R,\quad H_v^D=\sum_{ij}D_{ij}^v(\tau)\left(v_i-v_j\right)^2,\quad H_v^R=-\gamma\sum_{i}\left(u_iv_i-\alpha v_i^2\right), 
	\end{split}
\end{eqnarray}
where $H_{u,v}^D$ are the diffusion energies corresponding to the diffusion terms ${\Laplace}(\tau)u$ and ${\Laplace}(\tau)v$,  and $H_{u,v}^R$ denote the reaction energies corresponding to the reaction terms $f(u,v)$ and $g(u,v)$ (Appendix \ref{App-B}). The intensive parts $D_{ij}^{u,v}$ of $H_{u,v}^D$ have the same structure as $\Gamma_{ij}^G$ of $H_1^G$ in Eq. (\ref{discrete-Hamiltonian}) except for  $\chi_{ij}^{u,v}$ in Eq. (\ref{chi-uv}), and the detailed information of $\Gamma_{ij}^G$ is also given in Appendix \ref{App-A}. The normalization factors $\bar \Gamma_{u,v}$ associated with $D_{ij}^{u,v}$ in Eq. (\ref{discrete-HuvD}) differ from $\bar \Gamma_G$ of $\Gamma_{ij}^G$, because $\bar \Gamma_{u,v}$ are evaluated using $\chi_{ij}^{u,v}(\tau^{\rm iso})$. 
It should be noted that Finsler metric $g_{ab}^{u,v}(\tau)$ can also be incorporated into $H_{u,v}^R$ through their continuous representations,  $H_{u}^R\!=\!-\int \sqrt{g^{u}(\tau)}d^2x\left(u^2\!-\!\frac{u^4}{2}\!-\!uv\right)$ and $H_v^R\!=\!-\gamma\int \sqrt{g^{v}(\tau)}d^2x\left(uv\!-\!\alpha v^2\right)$, where $g^{u,v}(\tau)\!=\!\det{g_{ab}^{u,v}}$. However, such a modification was found to have no significant effect in simulations of fluctuating membranes \cite{Diguet-etal-PRE2024}. For this reason, in the non-fluctuating models we introduce $g_{ab}^{u,v}(\tau)$  only in the diffusion energies  $H_{u,v}^D$ [Eq. (\ref{discrete-Hu-Hv})] and in the diffusion terms of the RD equations (\ref{FN-eq}).

In the main text, the simulation results on the fixed lattices of $\varepsilon_R\!=\! 0$ (Fig. \ref{fig-2}(c)) will be presented. For this reason, the integration $\int_{\varepsilon_R} \Pi_{i}d\vec{r}_i$  inside $\varepsilon_R$ for the positional degrees of freedom (Fig. \ref{fig-2}) is not included in $Z_{\rm fix}$ in Eq. (\ref{part-funct}). The equivalence in the simulation results, as indicated in (\ref{nontrivila-phys-relation}), at small non-zero $\varepsilon_R$, corresponding to $\lim_{\varepsilon_R\to 0}\left.\mathcal{P}{hys}\right|_{\varepsilon_R}$, and at $\varepsilon_R\!=\! 0$, corresponding to $\left.{\rm Phys}\right|_{\varepsilon_R= 0}$, is shown in the supplementary material. Thus, the vertex position $\vec{r}$ is fixed in MC, nevertheless,  $\vec{r}$ as well as $\vec{\tau}$ is included in $H(\vec{r},\vec{\tau})$  in Eqs. (\ref{part-funct}) and (\ref{discrete-Hamiltonian}). This is because  the Finsler metric is defined using $\vec{r}$ as well as $\vec{\tau}$.

\subsection{Discrete RD equations and hybrid numerical technique \label{Discrete-RD-equation}}
	 To update the variables $u$ and $v$, a hybrid simulation technique is used. 
The discrete form of  RD equation in Eq. (\ref{FN-eq}) is 
\begin{eqnarray}
	\label{discrete-t-iterations}
	\begin{split}
		&u_{i}(t+{\Delta} t)\leftarrow u_{i}(t) +{\Delta} t \left(D_u{\Laplace}u_{i}(t) +f(u_{i}(t),v_{i}(t)) \right),\\
		&v_{i}(t+{\Delta} t)\leftarrow v_{i}(t) +{\Delta} t \left(D_v{\Laplace}v_{i}(t) +g(u_{i}(t),v_{i}(t))\right), \;(i=1,\cdots,N). 
	\end{split}
\end{eqnarray}
The discrete Laplacian $\Laplace u_i$ is given by Eq. (\ref{discrete-Laplace-App}).

The simulation procedure is as follows:
\begin{enumerate}
	\item[(i)] Initial values of $\{\vec{\tau}_i\}(\in S^{-1})$ \cite{note-1}, $u, v$ ($\in [-0.5,0.5]$) are randomly generated.
	\vspace{-2mm}
	\item[(ii)] 	One Monte Carlo sweep is performed to update 
	the variables $\vec{\tau}_i, (i\!=\!1,\cdots,N)$ using $H$ in Eq. (\ref{discrete-Hamiltonian}).
	The new variable $\vec{\tau}_i({\rm new})$ is accepted with the probability ${\rm Max}[1,\exp (-\delta H)]$ with  $\delta H\!=\!H(\vec{\tau}_i({\rm new}))\!-\!H(\vec{\tau}_i({\rm old}))$ \cite{Metropolis-JCP-1953,Landau-PRB1976}.
	\vspace{-2mm}
	\item[(iii)] The discrete time evolution of Eq. (\ref{discrete-t-iterations}) is iterated once with   ${\Delta} t \!=\!0.001$.
	\item[(iv)] Steps (ii) and (iii) are repeated $n_{\rm MC}$ times, where $n_{\rm MC}$ is suitably large. We assume
	\begin{eqnarray}
		\label{nMC-assumed}
		\begin{split}
			n_{\rm MC}=5\times 10^5 
		\end{split}
	\end{eqnarray}
	in the simulations for 2D square ($N\!=\!3600$) and triangular ($N\!=\!2900$) lattices, and 3D cubic lattice ($N\!=\!9600$).
	\item[(v)] Step (iii) is repeated under the final configurations of $\vec{\tau}_i, (i\!=\!1,\cdots,N)$ produced in (ii) until 
	the convergent criteria are satisfied:
	\begin{eqnarray}
		\begin{split}
			\label{convergence-FG}
			&{\rm Max}\left\{|u_{i}(t\!+\!\Delta t)\!-\!u_{i}(t)|\right\} \!<\! \varepsilon, \\
			&{\rm Max}\left\{|v_{i}(t\!+\!\Delta t)\!-\!v_{i}(t)|\right\} \!<\! \varepsilon,\quad (1\!\leq\!i\leq\!N),\\
			& \varepsilon = 1\!\times\! 10^{-7},\quad \Delta t\!=\!0.001. 
		\end{split}
	\end{eqnarray}
	\item[(vi)] Steps (i)--(v) are repeated 	$n_{\rm itr}$ times to calculate the mean values of physical quantities $Q(u,v,\vec{\tau})$ using the convergent configurations $\{u,v,\vec{\tau}\}_i, (i=1,\cdots,n_{\rm itr})$ obtained in step (iv) such that
	\begin{eqnarray}
	\label{mean-value}
	\langle Q\rangle= \frac{1}{n_{\rm itr}}\sum_{i=1}^{n_{\rm itr}} Q(\{u,v,\vec{\tau}\}_i). 
\end{eqnarray}
 We assume
	\begin{eqnarray}
		\label{nitration}
		n_{\rm itr}=200 \; ({\rm 2D}), \quad n_{\rm itr}=100 \; ({\rm 3D})
	\end{eqnarray}
	for the simulations of the 2D lattices and the 3D cubic lattice. 
	The initial configurations of the variables $u$, $v$ and $\vec{\tau}$ are randomly fixed using uniform random numbers such that $u,v \in \{-0.5,0.5\}$ and $\vec{\tau} \in S^1$: unit circle. 
\end{enumerate}

It should be noted that $n_{MC}$ is sufficiently large  in step (iii) for the convergence of $\vec{\tau}$; increasing $n_{MC}$ does not affect the results. Furthermore,  the TPs generated in step (iii) are almost identical to the convergent ones, even though the convergence criteria written in step (iv) are not met during the MC update of $\vec{\tau}$. This implies that the MC update of $\vec{\tau}$ significantly alters the TPs. Meanwhile, the time step of  the RD equations, ${\Delta}t$, alters the fine structures of $u$ and $v$, which are not visible in the TPs. Therefore, we consider  the change in the TPs caused by the RD equations to be negligible compared to the change caused by the MC update of $\vec{\tau}$.

	The lattice spacing $a$ in Eqs. (\ref{side-length-2D}) and (\ref{side-length-3D}) are discussed in Appendix \ref{App-D}.

\subsection{Mechanism for anisotropic Turing patterns \label{anisotropic-TP}}
In this subsection, we briefly describe the physical mechanism of anisotropic, or direction-dependent, diffusion that leads to the formation of anisotropic TPs. We use the square lattice to illustrate this process. Two different processes are involved in the formation of anisotropic TPs: 
\begin {enumerate}
\item[(i)] Lattice deformations align the orientation of IDOF $\vec{\tau}$
\item[(ii)] The aligned  $\vec{\tau}$ makes diffusion coefficients $D_{ij}^{u,v}$ direction-dependent  
\end{enumerate}
 The first process (i) is involved in the MC updates of $\vec{\tau}$, whereas the second process (ii) is involved in both the MC updates of $\vec{\tau}$ and the discrete-time updates in Eq.~(\ref{discrete-t-iterations}).  
In the process (i), the alignment of $\vec{\tau}$ is primarily caused by $H_1\!+\!\lambda H_\tau\!=\!\sum_{ij}\Gamma_{ij}^G(\tau)\ell_{ij}^2\!+\!\lambda H_\tau$.  If the lattice is deformed along the $x$ axis, then $\ell_{ij}>\ell_{kl}$ for all bonds $ij$ ($kl$) along the $x$ ($y$) direction. As a consequence,  the intensive part of $H_1$ satisfies the inequality $\Gamma_{ij}^G(\tau)<\Gamma_{kl}^G(\tau)$ due to the energy minimization principle. This condition $\Gamma_{ij}^G(\tau)<\Gamma_{kl}^G(\tau)$ suppresses the increase in  $\Gamma_{ij}^G(\tau)\ell_{ij}^2$  along the $x$ direction and the decrease in $\Gamma_{kl}^G(\tau)\ell_{kl}^2$  along the $y$ direction. The expression of $\Gamma_{i1}^G(\tau)$ for bond $i1$ (Fig.\ref{fig-4}(b)) is composed of the terms of the form  $\frac{\chi_{i1}^{G}}{\chi_{kl}^{G}}$, as shown in Eq. (\ref{discrete-H1-bonds-main}),  where bonds $kl$ are along the $y$ direction. Therefore, a small value of  $\Gamma_{i1}^G(\tau)$  implies  $\chi_{i1}^G<{\chi_{kl}^{G}}$. This implies that $\vec{\tau}_i$ is parallel to $\vec{e}_{i1}$ in the case of model 1,  because of the definition $\chi_{i1}^{G}\!=\!\sqrt{1-(\vec{\tau}_i\cdot\vec{e}_{i1})^2}+\chi_0$.  
Once $\vec{\tau}$ is aligned along the $x$ direction in the process (i), the aligned $\vec{\tau}$ makes $D_{ij}^{u,v}$ direction-dependent in the process (ii). Since $\vec{\tau}$ is parallel to the $x$ axis, $|\vec{\tau}_i\cdot\vec{e}_{ij}|$  becomes larger for bond $ij$ parallel to the $x$ axis compared to that for the perpendicular bonds. Therefore, $\chi_{ij}^{u}(=\!|\vec{\tau}_i\cdot\vec{e}_{ij}|\!+\!\chi_0)$ in Eq. (\ref{chi-uv}) is larger (smaller) for bond $ij$ parallel (perpendicular) to the $x$ axis. In contrast,  $\chi_{ij}^{v}(=\!	\sqrt{1-(\vec{\tau}_i\cdot\vec{e}_{ij})^2}\!+\!\chi_0)$  is smaller (larger) for bond $ij$ parallel (perpendicular) to the $x$ axis. These make $D_{ij}^{u,v}$ direction-dependent because $D_{ij}^{u,v}$ are defined as  $\frac{\chi_{ij}^{u,v}}{\chi_{kl}^{u,v}}$, analogous to the definition of $\Gamma_{ij}^G$ in Eq. (\ref{discrete-H1-bonds-main}).

\subsection{Mechanical anisotropy in 3D cubes \label{3D-cubes}}
\begin{figure}[h!]
	\centering{}\includegraphics[width=10.0cm]{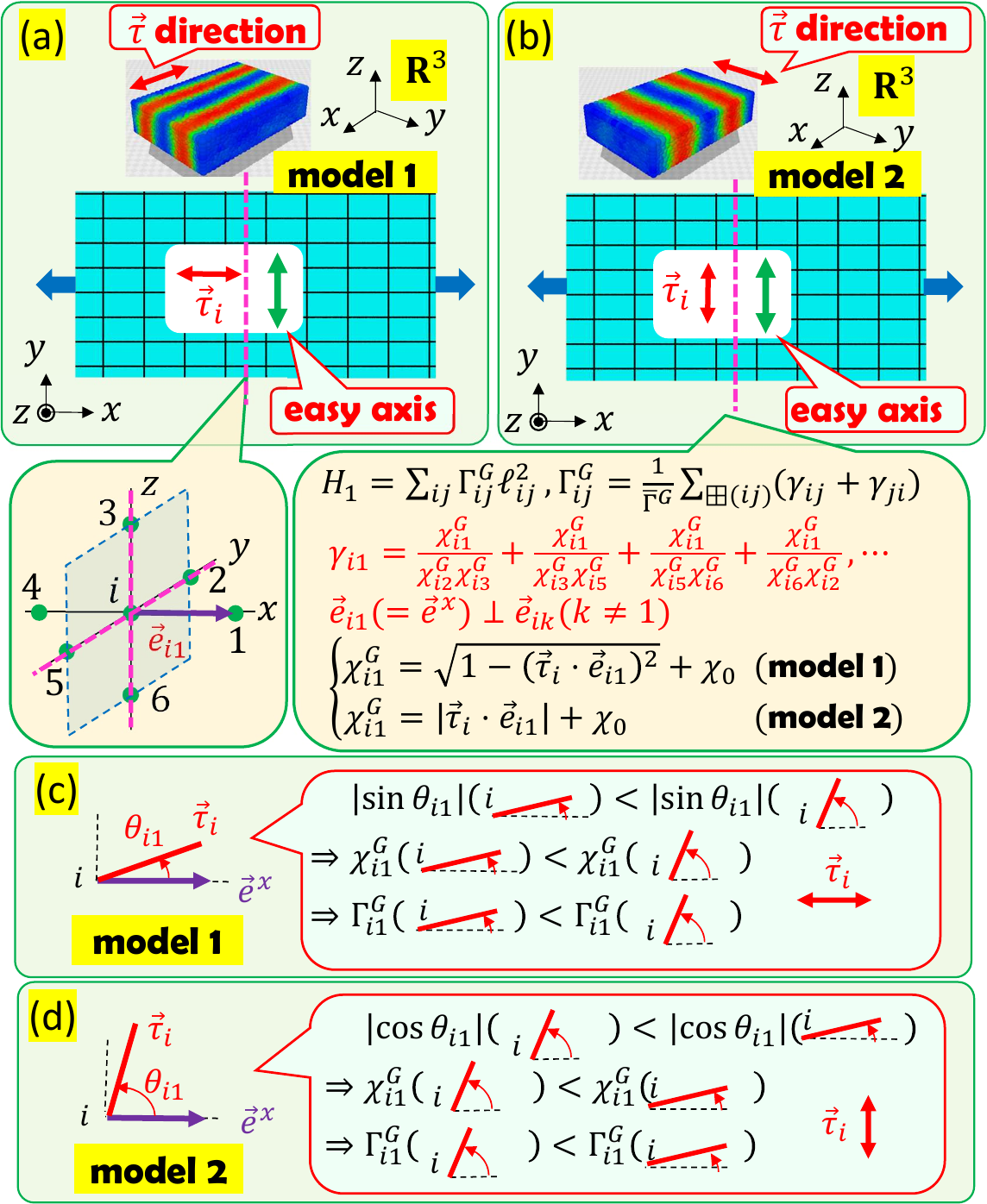}
	\caption { The TP direction is parallel to the $\vec{\tau}$ direction on the lattice extended along the $x$ direction of (a) model 1 and (b) model 2.  The magnitude of the unit Finsler length $\chi_{i1}^G$ along bond $i1$ depends on the model, thereby determining the $\vec{\tau}$ direction and the coupling constant $\Gamma_{i1}^G$, as shown in (c) and (d). Details of $\Gamma_{i1}^G$ in $H_1$ for the 3D cubic model are given in Eqs. (\ref{discrete-H1-tetra-3D}), (\ref{discrete-H1-coeff-3D}) and (\ref{discrete-H1-bond-1}).
		\label{fig-5} }
\end{figure}
As mentioned in the introduction, thin, rigid plate can be considered a 2D material. Therefore, it is reasonable to expect no difference in the mechanical anisotropy between the 3D cubic, thin plate and the 2D plates. In this subsection, we  demonstrate that the mechanical anisotropy in 3D rigid plates deformed oblong along the $x$-direction is equivalent to  that observed in 2D soft surfaces in Refs. \cite{Diguet-etal-PRE2024,FKato-SM2025}. This similarity is non-trivial because of the differences in the vertex mobility as well as in the material dimension. As stated at the beginning of Appendix  \ref{App-A-cubic}, the expression of $\Gamma_{ij}^G$ of the 3D cubic lattice is entirely distinct from that of the 2D lattices. Therefore, the question remains unanswered as to whether the mechanical anisotropy exhibited by 3D rigid plates can be intuitively understood as equivalent to that of 2D plates reported in \cite{FKato-SM2025}. 

Subsequent discussions reveal that the mechanical anisotropy exhibited by 3D cubic plates is analogous to that observed in 2D plates. 
Figures \ref{fig-5}(a), (b) illustrate deformed lattices with TPs of models 1 and 2 and the assumed directions of $\vec{\tau}$. The  Hamiltonian $H_1$ is written on the figures (see Appendix \ref{App-A-cubic} for detailed information of the 3D Hamiltonian).  The intensive part of the energy $\Gamma_{ij}^G$ on bond $i1$ along the $x$ direction depends on the unit Finsler length $\chi_{i1}^G$ in the numerator and $\chi_{ik}, (k\!=\!2,3,5,6)$ in the denominator. Due to this structure of  $\Gamma_{ij}^G$, we find that $\Gamma_{i1}^G$ of bond $i1$ along the $x$ direction becomes small (large) in model 1 (model 2) when $\vec{\tau}_i$ aligns along the $x$ ($y$) direction, as shown in Fig. \ref{fig-5}(c) (Fig. \ref{fig-5}(d)). It is important to note that the behaviors of  $\Gamma_{i1}^G$ are due to the increase of the extensive part $\ell_{i1}^2$ in the tensile energy $\Gamma_{i1}^G\ell_{i1}^2$. In this instance, the increase in bond length during deformation is more pronounced than in fluctuating lattices.  These behaviors in $\Gamma_{i1}^G$ are analogous to the 2D case in Ref. \cite{FKato-SM2025}. Thus, in model 1 of Fig. \ref{fig-5}(c), the directional energy localization from the $y$ axis to the $x$ axis is expected. This increase in energy along the $x$ axis makes the $y$ axis the easy axis for tensile deformation, as shown in Fig. \ref{fig-5}(a). The same discussion applies to the easy axis in model 2 as shown in Fig. \ref{fig-5}(b).

\section{Results \label{results}}
\subsection{Snapshots of Turing patterns on 2D and 3D plates\label{snapshot}}
\begin{figure}[h!]
	\centering{}\includegraphics[width=10.0cm]{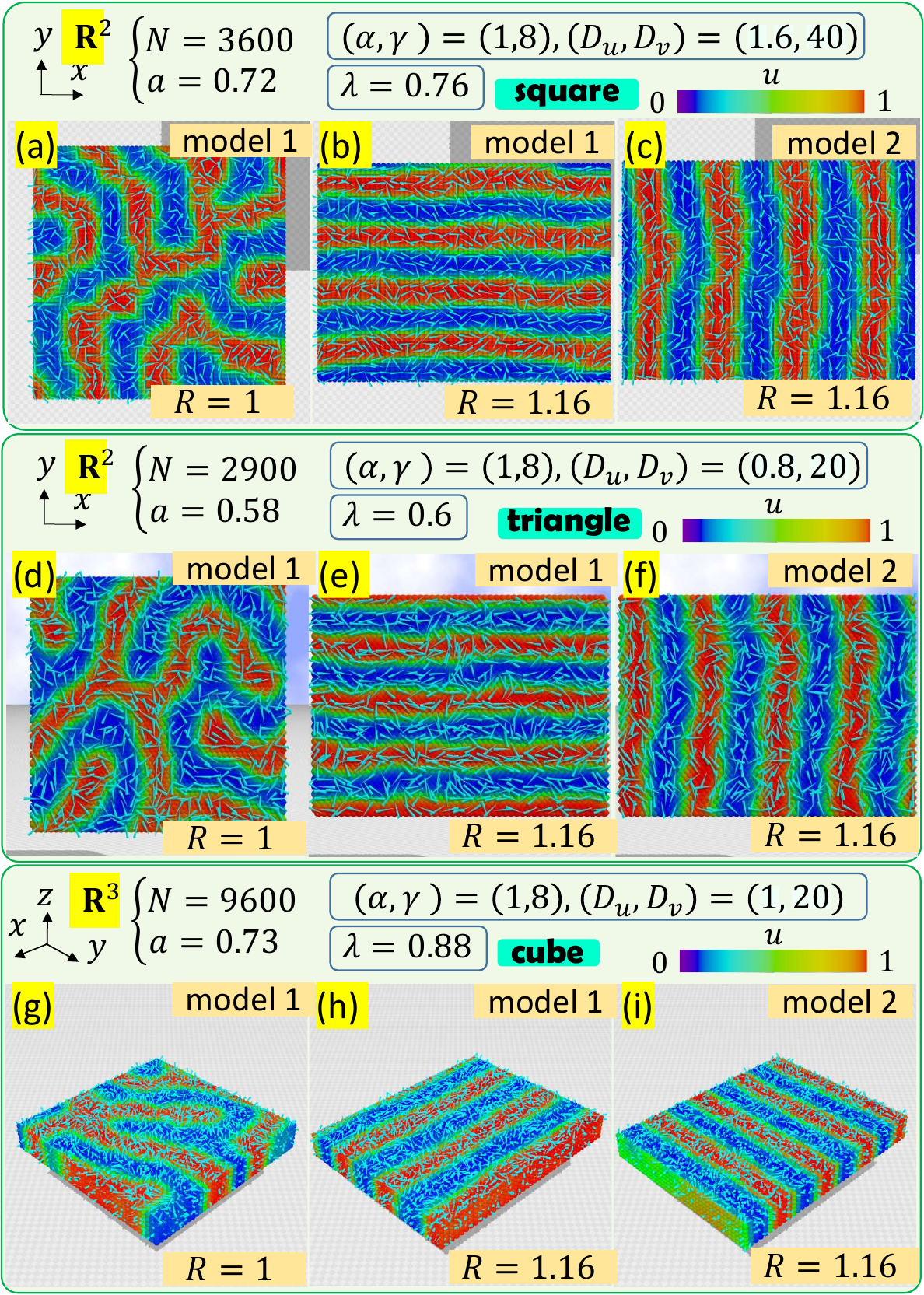}
	\caption {Snapshots of TPs on (a), (b) and (c) 2D square lattices, (d), (e) and (f) 2D triangular lattices, and (g), (h) and (i) 3D cubic lattices. The parameters assumed in the simulations are shown in the plots.  The small cones plotted at every second vertex represent the IDOF $\vec{\tau}$. The $\vec{\tau}$ direction is nearly parallel (perpendicular) to that of TP in model 1 (model 2).
		\label{fig-6}}
\end{figure}
Snapshots of the TPs are shown in Figs. \ref{fig-6}(a)--(c), \ref{fig-6}(d)--(f) and  \ref{fig-6}(g)--(i), which are obtained on 2D square and triangular lattices and the 3D cubic lattice, respectively. The assumed parameters are shown in the figures. Small cones indicate the stress directions, represented by $\vec{\tau}$, which are plotted at the every second vertex. While the correlation between the $\vec{\tau}$ direction and that of the TP is not so strong,  the two directions are correlated on a weakly basis. 
This correlation strength is also referred as "anisotropy strength",  as described in section \ref{model-Hamiltonian}.  This  strength can be controlled by $\chi_0$ (Eqs. (\ref{chi-G}) and (\ref{chi-uv}) in Appendix \ref{App-A}). For smaller values of  $\chi_0$, the anisotropy strength  becomes stronger.  The assumed value is $\chi_0\!=\!1$, which is not so small. However, the value $\chi_0\!=\!1$ is sufficient for the strain control of the TP direction. With the assumed parameters,  the TP direction is isotropic for $R\!=\!1$ in model 1 and becomes parallel (perpendicular) to  the external tensile force direction, which is the $x$ axis for $R\!=\!1.16$ in model 1 (model 2). The isotropic nature of the TP diretion at $R\!=\!1$ can also be seen in model 2 (not plotted) and is independent of the models. These observations are consistent with those in the 2D models for the soft materials in Refs. \cite{Diguet-etal-PRE2024,FKato-SM2025}. 

\subsection{Direction-dependent coefficients and energies for Gaussian bond potential on 2D plates \label{direction-dependent-GPB}}
\begin{figure}[h!]
	\centering{}\includegraphics[width=11.0cm]{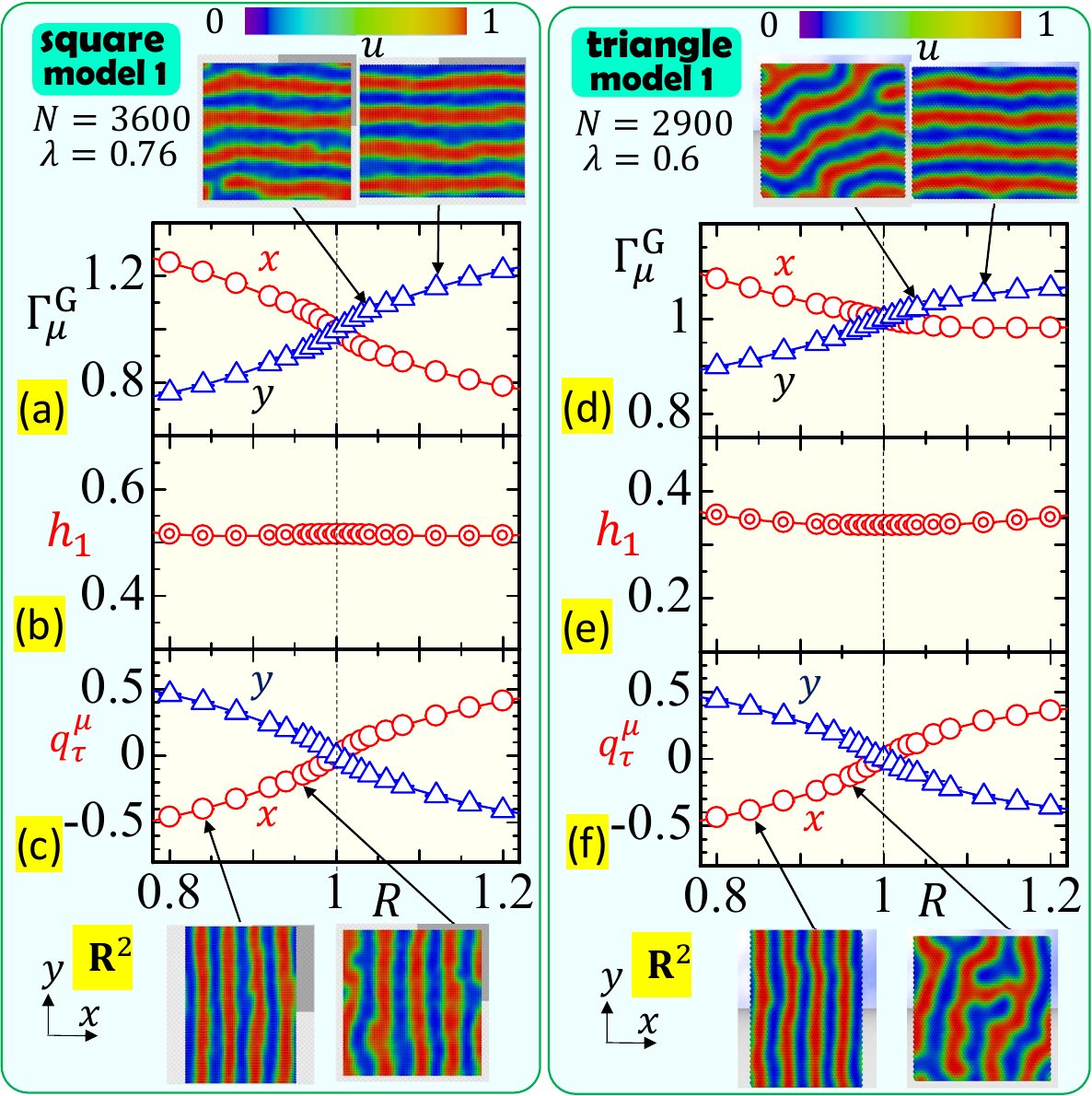}
	\caption {(a) $\Gamma_\mu^G$ vs. $R$, (b) $h_1(=\!H_1/N_B)$ vs. $R$, and (c) the order parameters $q_\tau^\mu$ vs. $R$ of the 2D square lattice model. The data of the 2D triangular lattice model are plotted in  (d), (e) and (f).   The slight asymmetry observed in data (d) and (f) under the $R\leftrightarrow 1/R$ transformation is due to the asymmetry of the triangular lattice structure along the $x$ and $y$ directions.
		\label{fig-7}}
\end{figure}
The results for the 2D plates are presented in this and the next subsection, while those for the 3D plates are presented in the following subsection.  The direction-dependent effective coefficients $\Gamma_\mu^G$ defined in Eq. (\ref{gamma-decomposition}) and the order parameters 
\begin{eqnarray}
	\label{order-prm}
		q_\tau^{\mu}=2\left(\frac{1}{N}\sum_i(\vec{\tau}_i\cdot\vec{e}^{\,\mu})^2-\frac{1}{2}\right), \quad (\mu=x,y) 
\end{eqnarray}
are plotted in Figs. \ref{fig-7}(a)--(f), where  $h_1\!=\!H_1/N_B$, where $N_B$ is the total number of bonds (see Eq. (\ref{energy-per-bond})).  Due to the normalization factor $\bar{\Gamma}_G$ in Eq. (\ref{discrete-H1-bond-1}), the directional components $\Gamma_\mu^G, (\mu=x,y)$  in Figs. \ref{fig-7}(a),(d) satisfy  $\Gamma_\mu^G\!\to\!1$ at $R\!\to\!1$,  where $\Gamma_\mu^G, (\mu=x,y)$ are given by Eq. (\ref{gamma-decomposition}).

The observed direction dependence in $\Gamma_\mu^G$ for an extension $R\!>\!1$ and compression $R\!<\!1$ of the plate along the $x$ direction is the same as that observed for membranes in Ref. \cite{FKato-SM2025}. Although the GBP can only vary in its intensive part $\Gamma_{ij}^G$ on the fixed lattices in this paper, it is interesting to observe the same responses to external mechanical stress as on the fluctuating lattices. On the fluctuating lattices, both the intensive and extensive parts of the GBP contribute to the tensile elasticity. On the fixed lattices, however, the tensile elasticity is only shared by the intensive part. The intensive part $\Gamma_{ij}^G$ varies depending on $\vec{\tau}$, and therefore, the tensile energy  $\Gamma_{ij}^G\ell_{ij}^2$ of bond $ij$ becomes dependent on the magnitude of  $\ell_{ij}^2$ even when $\ell_{ij}^2$ is fixed. 

The order parameters $q_\tau^{\mu}$ plotted in Figs. \ref{fig-7}(c),(f) have values $|q_\tau^{\mu}|\!\!<\! 0.5$, even though the range is $-1\!\leq\!q_\tau^{\mu}\!\leq\!1$ from Eq. (\ref{order-prm}). The reason for this relatively small  value  of $|q_\tau^{\mu}|$  for $R\!\to\!1.2$ and $R\!\to\!0.8$ is the large value of $\chi_0(=\!1)$, as mentioned above.

We briefly comment on the statistical errors in the data. Although standard deviations are shown as error bars, they are so small that they are barely visible in the figures. As described in Eq. (\ref{nitration}), all numerical results are obtained by averaging over  $n_{\rm itr}$ 
converged configurations, each generated from independently randomized initial values of $u$, $v$ and $\vec{\tau}$. Owing to the randomness of these initial conditions, the resulting converged configurations are not identical; in particular, the positions of the TPs vary from sample to sample. Nevertheless, the averaged physical quantities exhibit almost no variation, at least for the data presented in the main text. This is remarkable given that the total number of samples, $n_{\rm itr}$, is much smaller than that typically used in standard Monte Carlo simulations. The converged configurations correspond to solutions of the discrete RD equation, Eq. (\ref{discrete-t-iterations}), with uniformly aligned  $\vec{\tau}$ under lattice deformation characterized by $R(\not=\!1)$. The stochasticity affects only the global orientation of $\vec{\tau}$; however, because $\vec{\tau}$ remains uniformly aligned within each converged configuration, its contribution to fluctuations in the measured quantities is minimal, resulting in the extremely small error bars.

\subsection{Direction-dependent diffusion constants and energies on 2D plates \label{direction-dependent-quantity}}
\begin{figure}[h!]
	\centering{}\includegraphics[width=11.0cm]{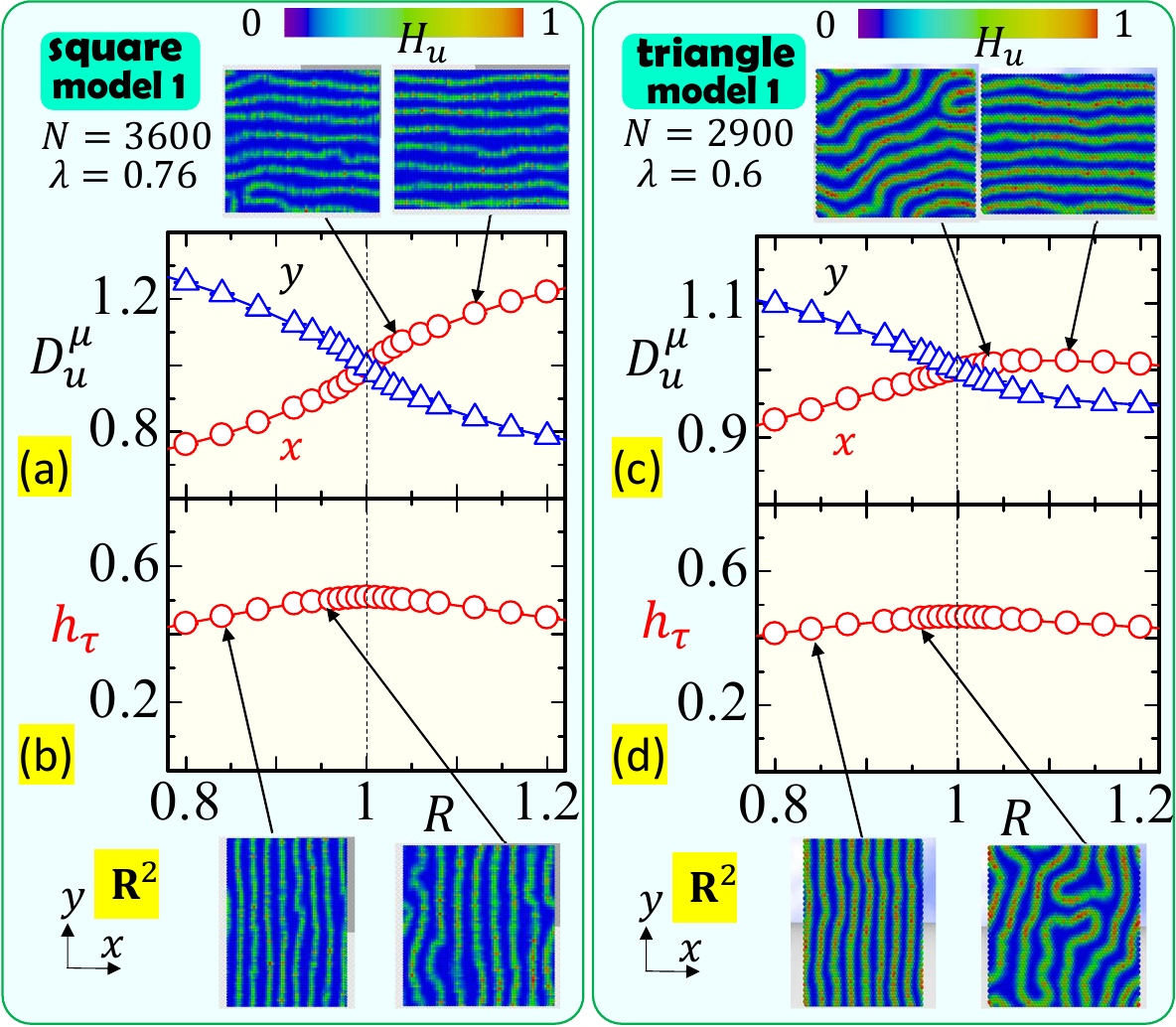}
	\caption {(a) $D_u^\mu$ vs. $R$ and (b) $h_\tau(=\!H_{\tau}/N_B)$ vs. $R$ of the 2D square lattice model, and (c), (d) the data of the 2D triangular lattice model.  
		\label{fig-8}}
\end{figure}
The direction-dependent diffusion coefficients are given by 
\begin{eqnarray}
\label{effective-diffusion-constants-phsi}
\begin{split}
D_{u,v}^{\mu}=\frac{1}{\sum_{ij}|\vec{e}_{ij}\cdot\vec{e}^{\,\mu}|}\sum_{ij}D_{ij}^{u,v}|\vec{e}_{ij}\cdot\vec{e}^{\,\mu}|, \quad (\mu=x,y), 
\end{split}
\end{eqnarray}
where $D_{ij}^{u,v}$ is defined by Eq. (\ref{discrete-HuvD}).
These quantities and $h_\tau\!=\!H_\tau/N_B$ are plotted in Figs. \ref{fig-8}(a)--(f). The coefficients $D_{u,v}^\mu, (\mu\!=\!x,y)$  are distinctively split into two different values when $R$ deviates from $R\!=\!1$ to $R\!\not=\!1$ (Figs. \ref{fig-8}(a), (c)).

\subsection{Results of 3D plate \label{direction-control}}
\begin{figure}[h!]
	\centering{}\includegraphics[width=11.0cm]{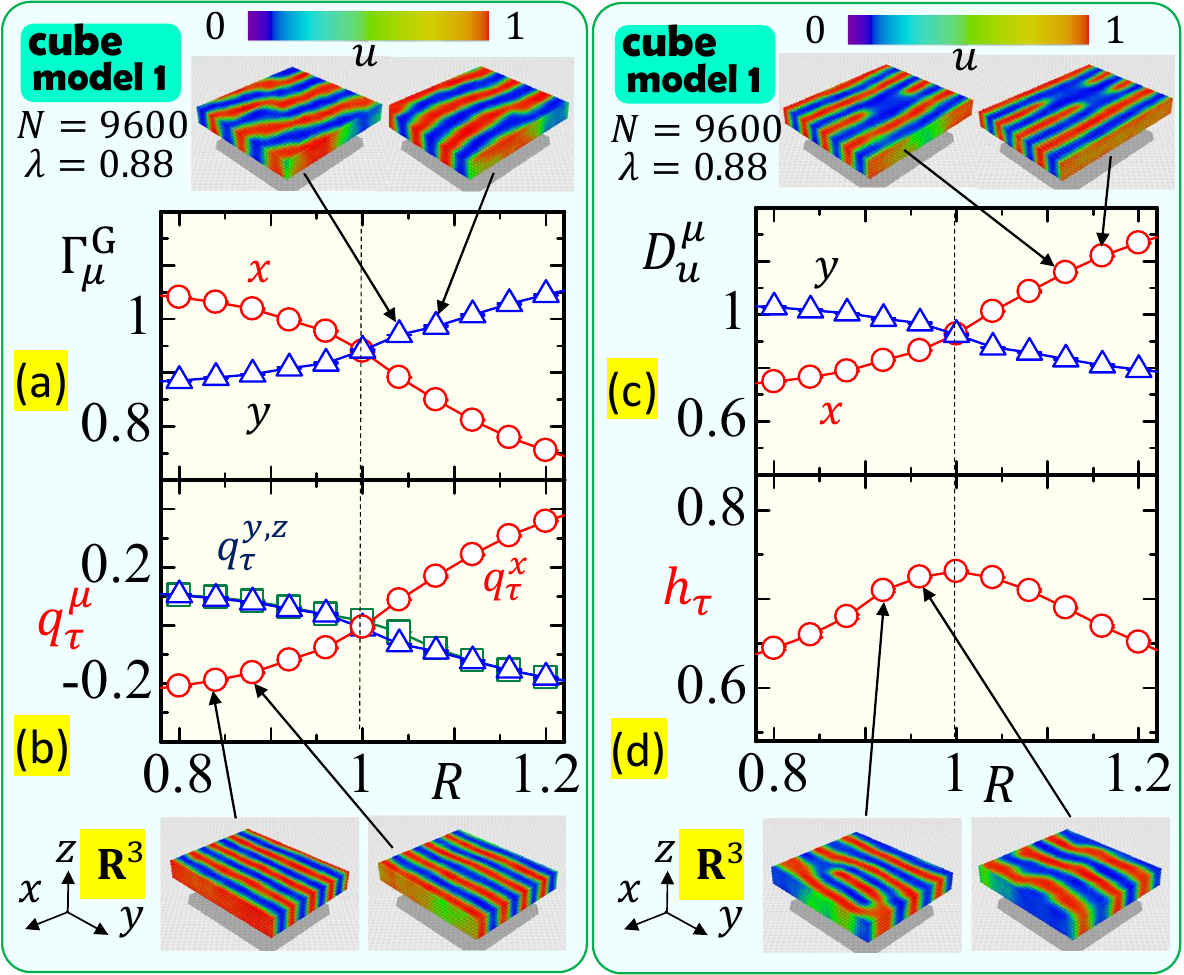}
	\caption {(a) $\Gamma_\mu^G$ vs. $R$, (b) $q_\tau^\mu$ vs. $R$ of the 3D cube lattice model of size $N\!=\!9600$.  (c) $D_u^\mu$ vs. $R$,  and (d) $h_\tau(=\!H_{\tau}/N_B)$ vs. $R$ of the same model.    Asymmetry with respect to $R\!\leftrightarrow\!1/R$ observed in data is due to the assymetry of $L_x$ and $L_y$ of 3D model in Eq. (\ref{side-length-3D}) in contrast to those of 2D models in Eq. (\ref{side-length-2D}).
		\label{fig-9}}
\end{figure}
This section presents results that are not trivial, in the sense that Eq. (\ref{nontrivila-phys-relation})  is assumed in the calculation. The discrete Hamiltonian defined on the 3D cubic lattice is presented in Appendix \ref{App-A-cubic}. 

First, we show the results, which are independent of the assumption in Eq. (\ref{nontrivila-phys-relation}). The results of the directional components $\Gamma_\mu^G$  of GBP, and the order parameters $q_\tau^\mu(=(3/2)\sum_i[(\vec{\tau}_i\cdot\vec{e}^{\,\mu})^2/N\!-\!(1/3)])$ obtained on 3D cube are plotted in Figs. \ref{fig-9}(a),(b). It is observed that the behavior of the data with respect to the $R$ variation is nearly identical to that of the 2D plates in Fig. \ref{fig-7}. Since the $x$ and $y$ components can only be compared to the 2D results, no $z$ component is plotted. The quantities corresponding to the diffusion energy coefficient and $h_\tau$ plotted in Figs. \ref{fig-9}(c),(d) are also consistent with the 2D data in Fig. \ref{fig-8}.

\begin{figure}[h!]
	\centering{}\includegraphics[width=11.0cm]{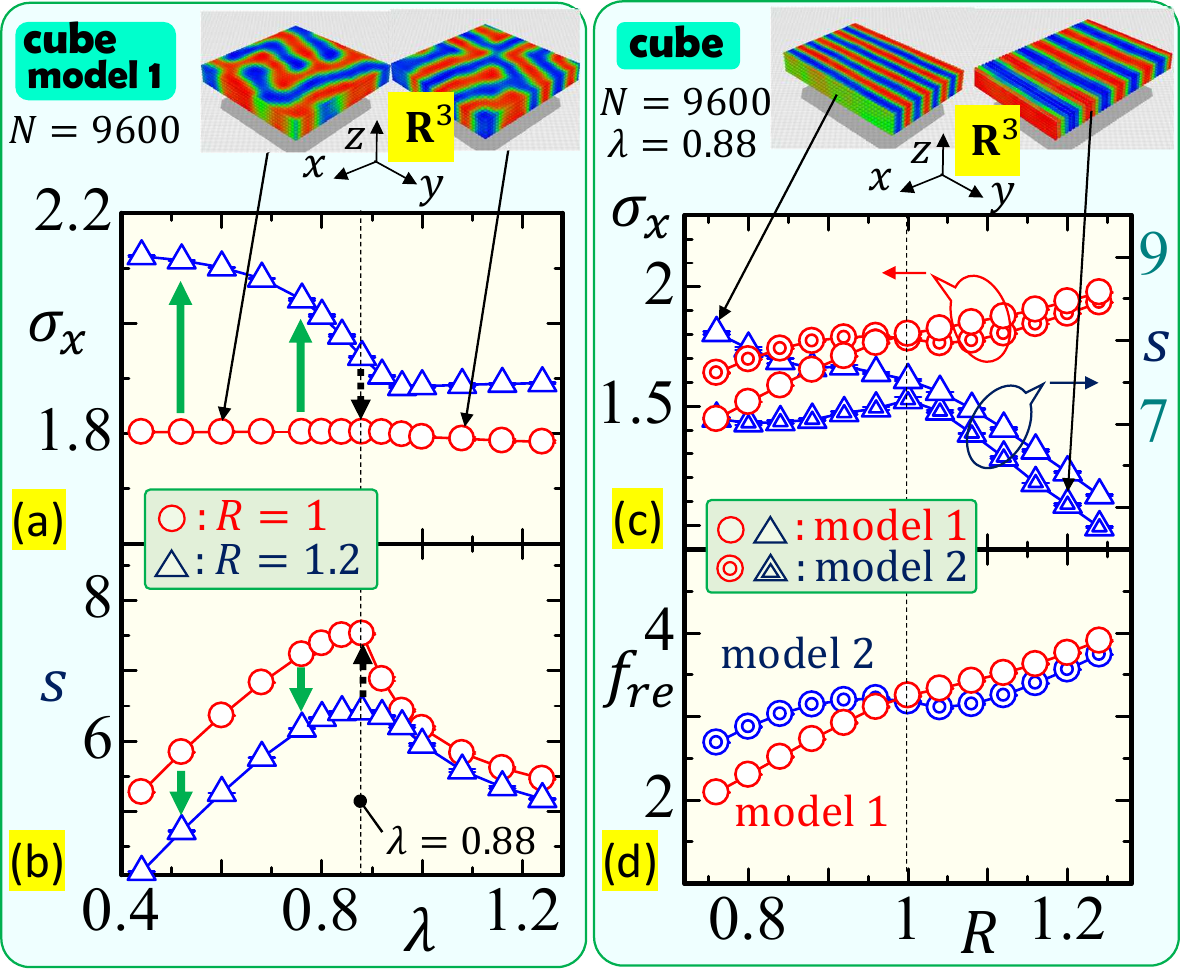}
	\caption {(a) $\sigma_x$ vs. $\lambda$ and (b) $s$ vs. $\lambda$ of 3D plate model 1 for $R\!=\!1 (\textcolor{red}{\bigcirc})$ and $R\!=\!1.2(\textcolor{blue}{\triangle})$.  The entropy density $s$ has a peak at $\lambda_c\!\simeq\!0.88$ for both $R\!=\!1$ and $R\!=\!1.2$. (c) $\sigma_x$ vs. $R$ and $s$ vs. $R$, and (d) free energy density $f_{\rm re}$, of models 1 and 2.
		\label{fig-10}} 
\end{figure}
Now, we presents the results obtained under the assumption in Eq. (\ref{nontrivila-phys-relation}). The formulas for stress $\sigma_x$, entropy $s$, and free energy $f_{\rm re}$ are presented in sections IIIA and IIIC of  the supplementary material,  which primarily presents data obtained from 2D plates. The 3D extensions are straight forward, and we have
\begin{eqnarray}
	\label{stress-entropy-free}
	\begin{split}
	&\sigma_x=\sqrt{\frac{2\Gamma_x^G H_1^x}{V}}, \\
	&s(V)=\frac{H_1+\lambda H_\tau+\left(H_u-\frac{1}{\gamma}H_v\right)-2\Gamma_x^GH_1^x}{V},\\
	&f_{\rm re}=\frac{2\Gamma_x^G H_1^x}{V},
	\end{split}
\end{eqnarray}
where $V$ denotes the volume.
The stress $\sigma_x$ vs. $\lambda$ obtained on the 3D lattice for $R\!=\!1$ remains almost constant (Fig. \ref{fig-10}(a)), while the entropy $s$ has a peak at $\lambda\!\simeq\!0.88$ (Fig. \ref{fig-10}(b)). Notably, this peak position remains unaltered  when $R$ is changed to $R\!=\!1.2$. 
On the right-hand side of the expression for $s$, the terms $H_u^R$ and $H_v^R$ are excluded from $H_u\!-H_v/\gamma$ because they are not treated as internal energies in the MC update of $\vec{\tau}$. Even if $H_u^R$ and $H_v^R$ are included in the calculation of $s$, the shape of the entropy curve remains essentially unchanged. The maximum of the entropy at $\lambda\!\simeq\!0.88 (=\!\lambda_c)$ indicates an equilibrium stable state. The fact that the peak position is independent of $R$ suggests that this state remains stable under the lattice deformation.  It is therefore reasonable to expect that the stability point $\lambda_c$ is robust against external mechanical perturbations, including variations in $\chi_0$ \cite{FKato-SM2025}, although such variations are not examined in the present study. This robustness can be understood from the role of $\lambda$ as the coupling constant of the correlation energy
 $H_\tau$ in Eq. (\ref{discrete-Hamiltonian}). The correlation of $\vec{\tau}$ depends only on intrinsic lattice properties, such as the bond connectivity, and is independent of the external mechanical state of the lattice. By contrast, the alignment direction  of  $\vec{\tau}$ is not controlled by  $\lambda$ through $H_\tau$; rather, it is governed solely by the lattice deformation.

The stress $\sigma_x$ at $R\!=\!1.2$ in Fig. \ref{fig-10}(a) rapidly decreases with increasing $\lambda$ at the peak position $\lambda\!\simeq\!0.88$ of $s$. However, the relations $\sigma_x(R\!=\!1.2)\!>\!\sigma_x(R\!=\!1)$ and $s(R\!=\!1.2)\!<\!s(R\!=\!1)$ at each $\lambda$ remain unchanged. These relations indicate that the plate extension from $R\!=\!1$ to $R\!=\!1.2$ decreases the entropy (Fig. \ref{fig-10}(b)) and  is accompanied by an increase in $\sigma_x$ (Fig. \ref{fig-10}(a)) as indicated by the solid arrows. The converse phenomenon indicated by the dashed arrows at $\lambda\!=\!0.88$ corresponds to the stress relaxation. This phenomenon is expected when an external force  is released from a deformed lattice characterized by $R \!=\! 1.2$.    

The stress $\sigma_x$ and entropy $s$ exhibit a reciprocal relationship with the ratio $R$ within the extension region, defined as $R>1$. This relationship is illustrated in Fig. \ref{fig-10}(c) for both models 1 and 2. These behaviors are consistent with the results of 2D plates presented in the supplementary material. This response is consistent with the entropy elasticity model, in which a tensile elasticity is caused by an entropy decrement.  The free energy density $f_{\rm re}$ vs. $R$ in Fig. \ref{fig-10}(d) is similar to the behavior of $\sigma_x$ in Fig. \ref{fig-10}(c) in both models, as expected. Note that $\sigma_x$, $s$ and $f_{\rm re}$ in Figs. \ref{fig-10}(c), (d) increase with increasing $R$  including the region $R\!<\!1$. This behavior is attributed to the response to the plate extension along the positive $x$ direction.

\section{Concluding remarks \label{conclusion}}
This paper explores Turing patterns (TPs) on non-fluctuating or fixed lattices, using a Finsler geometry (FG) model with a hybrid simulation technique combining Monte Carlo methods and discrete Turing equations.  The internal degree of freedom (IDOF) plays a crucial role in the anisotropy of elasticity and TPs, and the tensile stress and entropy are evaluated as in models on fluctuating lattices for membranes.

Hybrid simulations are executed on three distinct lattices: two-dimensional (2D) regular square and triangular lattices, as well as  a three-dimensional (3D) cubic plate. In contradistinction to elastic membranes, the vertex positions are fixed, and no explicit bond elasticity is assumed.  Despite the lack of bond elasticity caused by the fixed vertices in the Gaussian bond potential (GBP), the IDOF in the intensive part of the GBP responds to the lattice deformation  and determines the TP direction,  in a manner analogous to fluctuating lattices.  We find that the responses of TP direction to the lattice deformation of the fixed 2D and 3D lattices are nearly identical to those in reported data of fluctuating lattices for membranes.

Furthermore, the entropy exhibits a peak value at $\lambda\!=\!\lambda_c$, where $\lambda$ is the coefficient of the IDOF Hamiltonian. The peak position $\lambda_c$ of the entropy density is independent of the lattice deformation ratio $R$ in the 2D triangular and 3D cubic models. Since the IDOF is regarded as a stress direction, the maximal entropy condition indicates that the variation of IDOF at $\lambda_c$ corresponds to stress relaxation during the lattice deformation,  similar to fluctuating membranes.

	Additionally,  TPs are expected to be isotropic in the limits $R\!\to\!1$ and  $\chi_0\!\to\!\infty$, regardless of the values of $\chi_0(>\!0)$ and $R$, respectively. It would therefore be of interest  to construct a phase diagram in the $(R,\chi_0)$ plane and investigate whether a sharp phase boundary separates isotropic and anisotropic TPs.

\begin{acknowledgments}
This work was supported in part by Collaborative Research Project J25Ly01 of the Institute of Fluid Science (IFS), Tohoku University. Numerical simulations were performed under Project CP04JUN25 on the AFI-NITY supercomputer system at the Advanced Fluid Information Research Center, Institute of Fluid Science, Tohoku University. 
\end{acknowledgments}

\appendix

\section{Discretization of the Gaussian bond potential \label{App-A}}
\subsection{Discrete Finsler metric \label{discrete-Finsler-metric}}
We present how the discrete expression of $H_1$ in Eq. (\ref{discrete-Hamiltonian}) is obtained from the continuous Hamiltonian
\begin{eqnarray}\label{continuous-H1}
	H_1=\int \sqrt{g^G}d^2x g_G^{ab}\frac{\partial \vec{r}}{\partial x^a}\cdot\frac{\partial \vec{r}}{\partial x^b}, 
\end{eqnarray}
where $g_G^{ab}$ is the inverse of the Finsler metric $g_{ab}^G$ at vertex $i$, and it is given by (Fig. \ref{fig-A-1}(a))
\begin{eqnarray}
	\label{F-metric}
	\begin{split}
	&g_{ab}^G=\begin{pmatrix}
		(\chi_{i1}^G)^{-2} & 0\\
		0 & (\chi_{i2}^G)^{-2} 
	\end{pmatrix},\\
	 &\sqrt{g^G}=\sqrt{\det g_{ab}^G}=\frac{1
	}{\chi_{i1}^{G}\chi_{i2}^{G}}, \quad g^{ab}_G=(g_{ab}^G)^{-1}.
	\end{split}
\end{eqnarray}
\begin{figure}[h]
	\centering{}\includegraphics[width=11.0cm,clip]{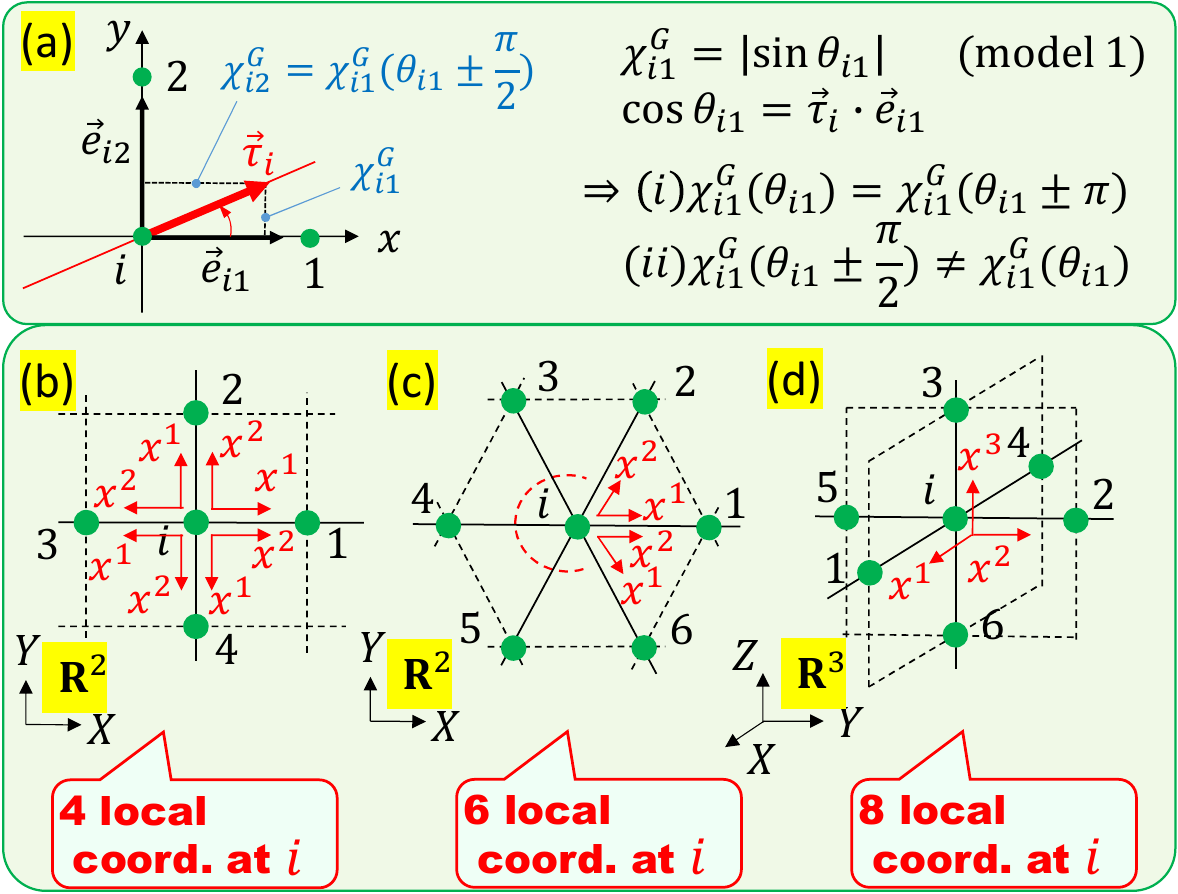}
	\caption{
		(a) An example of IDOF variabl $\vec{\tau}_i$ at vertex $i$ exhibiting (i) rotational symmetry  and (ii) anisotropy of $\chi_{i1}^G$ (Eq.(\ref{chi-G-i1})).   Lattice structures of (b) square, (c) triangular, and (d) cubic lattices.  (b) Four possible local coordinates at vertex $i$ of square lattice with four vertices $1, 2, 3$ and $4$, (c) six possible local coordinates at vertex $i$ of triangular lattice with vertices $1,2,\cdots, 6$. (d) Eight local coordinates at vertex $i$ of the cubic lattice with 6 vertices $1,\cdots,6$, where a local coordinate $(x^1,x^2,x^3)$ is plotted on the figure.  
		\label{fig-A-1}}
\end{figure}
The $\chi_{ij}^G, (j\!=\!1,2)$, defined by
\begin{eqnarray}
	\label{chi-G}
\chi_{ij}^G=\left\{ \begin{array}{@{\,}ll}
	\sqrt{1-(\vec{\tau}_i\cdot\vec{e}_{ij})^2}+\chi_0\quad  (\sin)\; ({\rm for\; model\; 1}) \\
	|\vec{\tau}_i\cdot\vec{e}_{ij}|+\chi_0\qquad\qquad  (\cos)\; ({\rm for\; model\; 2}) 
\end{array} \right.,\quad \chi_0=1,
\end{eqnarray}
represent the unit Finsler length from vertex $i$ along the local coordinate axes $x^j$  (Figs. \ref{fig-A-1}(a)--(d)), where $\vec{e}_{ij}$ is the unit vector from vertices $i$ to $j$. The interaction strength depends on $\chi_{ij}^G$, a factor that exerts a non-trivial influence on the intensive part of the discrete Hamiltonian. Consequently, the discrete expression, which depends on the lattice structure, is slightly complex compared with the case of Euclidean metric corresponding to $\chi_{ij}^G\!=\!1$.

In the case of model 1, we have the unit Finsler length $\chi_{i1}^G\!=\!|\sin \theta_{i1}|$ between vertices  $i$ and $j(=\!1)$  from Eq. (\ref{chi-G}), where $\theta_{i1}$ is the angle defined by $\cos\theta_{i1}\!=\!\vec{\tau}_{i1}\cdot \vec{e}_{i1}$ and  $\chi_0\!=\!0$ is assumed for simplicity on the square lattice (Fig. \ref{fig-A-1}(a)). This  $\chi_{i1}^G$ satisfies
\begin{eqnarray}
	\label{chi-G-i1}
\left\{ \begin{array}{@{\,}ll}
	({\rm i})  \chi_{i1}^G(\theta_{i1})= \chi_{i1}^G(\theta_{i1}\pm\pi)\\
	({\rm ii})  \chi_{i2}^G(\theta_{i1})=\chi_{i1}^G(\theta_{i1}\pm\frac{\pi}{2})\not=\chi_{i1}^G(\theta_{i1})
\end{array} \right..
\end{eqnarray}
The relation (i) signifies a symmetry under the rotation of $\pi$, while the relation (ii)  represents an anisotropy of interaction strength. Note that $\chi_{i1}^G(\theta_{i1})$ and $\chi_{i2}^G(\theta_{i1})$ in (ii)  correspond to the diagonal components of $g_{ab}^G$ in Eq. (\ref{F-metric}). Therefore, the anisotropy (ii) is due to the rotational asymmetry of $g_{ab}^G$. This anisotropy of $\chi_{i1}^G(\theta_{i1})$ depends on the direction of $\vec{\tau}_{i1}$, and is dynamically changeable, and $\chi_{i1}^G(\theta_{i1})$ attains its maximum at the $\vec{\tau}_{i1}$ configuration of $|\vec{\tau}_{i1}\cdot \vec{e}_{i1}|\!\to\! 1$ or equivalently $\theta_{i1}\!\to\!0,\pi$.

The unit Finsler lengths $\chi_{ij}^{u,v}$ for the diffusion interaction $H_{u,v}^D$, which will be given below,  are defined to be the cosine type and the sine type, respectively, such that
\begin{eqnarray}
	\label{chi-uv}
	\begin{split}
	&\chi_{ij}^u=|\vec{\tau}_i\cdot\vec{e}_{ij}|+\chi_0\quad  (\cos), \\
	&\chi_{ij}^v=	\sqrt{1-(\vec{\tau}_i\cdot\vec{e}_{ij})^2}+\chi_0\quad  (\sin),\\
	&\chi_0=1.
	\end{split}
\end{eqnarray}
These relations are used in combination with $\chi_{ij}^G$ in both models 1 and 2 as those defined in Eq. (\ref{chi-G}).

\subsection{Discrete Hamiltonian on square lattice \label{App-A-square}}
Replacing the integral with the sum over squares and the differential with the difference such that
\begin{eqnarray}
	\label{discretization}
	\begin{split}
		&\int \sqrt{g^G}d^3x\to\sum_\square \sqrt{g^G}\\
		&\frac{\partial \vec{r}_i}{\partial x^j}\to\vec{r}_j\!-\!\vec{r}_i,
	\end{split}
\end{eqnarray}
we obtain  $\sum_\square \sqrt{g^G} \left(g_G^{i1}\ell_{i1}^2\!+\!g_G^{i2}\ell_{i2}^2\right)$, where the local coordinate $(x^1,x^2)$ is assumed as shown in Fig. \ref{fig-A-1}(b), and  $\ell_{ij}\!=\!\|\vec{r}_j\!-\!\vec{r}_i\|$.

Using the expressions in Eq. (\ref{F-metric}), we have
\begin{eqnarray}
	\label{discrete-H1-org-square}
	\begin{split}
H_1&\to\sum_\square \frac{1
}{\chi_{i1}^{G}\chi_{i2}^{G}} \left((\chi_{i1}^G)^{2}\ell_{i1}^2+(\chi_{i2}^G)^2\ell_{i2}^2\right)\\
&=\sum_\square  \left(\frac{\chi_{i1}^{G}
}{\chi_{i2}^{G}}\ell_{i1}^2+\frac{\chi_{i2}^{G}
}{\chi_{i1}^{G}}\ell_{i2}^2\right).
	\end{split}
\end{eqnarray}
There are four possible local coordinates at vertex $i$ on the square lattice (Fig. \ref{fig-A-1}(b)). 
This summation convention for the local coordinates at vertex $i$ is equivalent to that for the four vertices of a square.  Therefore, summing over four possible expressions of $H_1$, which are obtained by replacing $(i1, i2)\!\to\!(i2, i3)$, $(i3, i4)$, $(i4, i1)$ in Eq. (\ref{discrete-H1-org-square}), we have
 $6(=\!2\!\times\! 3)$ additional terms such that
\begin{eqnarray}
	\label{discrete-H1-other}
	\begin{split}
		&\frac{\chi_{i2}^{G}
		}{\chi_{i3}^{G}}\ell_{i2}^2+\frac{\chi_{i3}^{G}
		}{\chi_{i2}^{G}}\ell_{i3}^2 \;
		+\;\frac{\chi_{i3}^{G}
		}{\chi_{i4}^{G}}\ell_{i3}^2+\frac{\chi_{i4}^{G}
		}{\chi_{i3}^{G}}\ell_{i4}^2 \;
		+\;\frac{\chi_{i4}^{G}
		}{\chi_{i1}^{G}}\ell_{i4}^2+\frac{\chi_{i1}^{G}
		}{\chi_{i4}^{G}}\ell_{i1}^2.
	\end{split}
\end{eqnarray}
Since the sum over squares $\sum_\square$ can be replaced by the sum over vertices  $\sum_i$, we have  
\begin{eqnarray}
	\label{discrete-H1-tetra}
	\begin{split}
		H_1=&
		\sum_i  \left(\gamma_{i1}^G\ell_{i1}^2+\gamma_{i2}^G\ell_{i2}^2+\gamma_{i3}^G\ell_{i3}^2+\gamma_{i4}^G\ell_{i4}^2\right)\\
		=&\sum_i\sum_{j(i)}\gamma_{ij}^G\ell_{ij}^2=\frac{1}{2}\sum_i\sum_{j(i)}\gamma_{ij}^G\ell_{ij}^2+\frac{1}{2}\sum_j\sum_{i(j)}\gamma_{ji}^G\ell_{ji}^2\\
		=&\frac{1}{2}\sum_i\sum_{j(i)}\left(\gamma_{ij}^G+\gamma_{ji}^G\right)\ell_{ij}^2
		=\sum_{ij}\left(\gamma_{ij}^G+\gamma_{ji}^G\right)\ell_{ij}^2\\
		=&\sum_{ij}\Gamma_{ij}^G\ell_{ij}^2,\quad \Gamma_{ij}^G=\gamma_{ij}^G+\gamma_{ji}^G,
	\end{split}
\end{eqnarray}
where $\gamma_{ij}^G\!\not=\!\gamma_{ji}^G$ in the final expression, and $\gamma_{ij}^G, (j=1,\cdots,4)$ are given by 
\begin{eqnarray}
	\label{discrete-H1-coeff}
	\begin{split}
		\gamma_{i1}^G=\frac{\chi_{i1}^{G}}{\chi_{i2}^{G}}+\frac{\chi_{i1}^{G}}{\chi_{i4}^{G}}, \; \gamma_{i2}^G=\frac{\chi_{i2}^{G}}{\chi_{i1}^{G}}+\frac{\chi_{i2}^{G}}{\chi_{i3}^{G}}, \;
		\gamma_{i3}^G=\frac{\chi_{i3}^{G}}{\chi_{i2}^{G}}+\frac{\chi_{i3}^{G}}{\chi_{i4}^{G}}, \; \gamma_{i4}^G=\frac{\chi_{i4}^{G}}{\chi_{i3}^{G}}+\frac{\chi_{i4}^{G}}{\chi_{i1}^{G}}, \quad (\square). 
	\end{split}
\end{eqnarray}
The sum of four terms in the first of Eq. (\ref{discrete-H1-tetra}) can be written by using  $\sum_{j(i)}$; the sum over vertices $j$ connected to $i$, in the second line.  
On the third line in Eq. (\ref{discrete-H1-tetra}), the relation $(1/2)\sum_i\sum_{j(i)}\!=\!\sum_{ij}$ is used, where $\sum_{ij}$ is the sum over bonds $ij$.  $\Gamma_{ij}^G(=\!\gamma_{ij}^G\!+\!\gamma_{ji}^G)$ is suitably normalized. This will be shown in the following subsection. 

We should note that the four terms in the first of Eq. (\ref{discrete-H1-tetra}) correspond to the bonds connected to vertex $i$, and the two terms of $\gamma_{ij}^G$ in Eq. (\ref{discrete-H1-coeff}) imply that bond $ij$ is shared by two squares. Periodic boundary condition is assumed, except the free boundary of the upper and lower surfaces of the 3D plate and the virtual boundary assumed in the calculation of stresses, which is shown in the supplementary material.

\subsection{Discrete Hamiltonian on triangular lattice \label{App-A-triangle}}
Using the local coordinate $(x^1,x^2)$ of the triangle $i12$ in Fig. \ref{fig-A-1}(c), we have
\begin{eqnarray}
	\label{discrete-H1-org-triangle}
	\begin{split}
		H_1&\to\sum_\triangle \frac{1
		}{\chi_{i1}^{G}\chi_{i2}^{G}} \left((\chi_{i1}^G)^{2}\ell_{i1}^2+(\chi_{i2}^G)^2\ell_{i2}^2\right)\\
		&=\sum_\triangle  \left(\frac{\chi_{i1}^{G}
		}{\chi_{i2}^{G}}\ell_{i1}^2+\frac{\chi_{i2}^{G}
		}{\chi_{i1}^{G}}\ell_{i2}^2\right),
	\end{split}
\end{eqnarray}
which is the same as that in Eq. (\ref{discrete-H1-org-square}).  Since there are three possible local coordinates on a triangle, by replacing $(i1,i2)$ with $(12,1i)$ and $(2i,21)$, and by including the additional terms
\begin{eqnarray}
\frac{\chi_{12}^{G}
}{\chi_{1i}^{G}}\ell_{12}^2+\frac{\chi_{1i}^{G}
}{\chi_{12}^{G}}\ell_{1i}^2 +\frac{\chi_{2i}^{G}
}{\chi_{21}^{G}}\ell_{2i}^2+\frac{\chi_{21}^{G}
}{\chi_{2i}^{G}}\ell_{21}^2,
\end{eqnarray}
we have
\begin{eqnarray}
	\label{discrete-H1-triangle}
	\begin{split}
		&H_1=
		\sum_\triangle  \left(\gamma_{i1}^G\ell_{i1}^2+\gamma_{12}^G\ell_{12}^2+\gamma_{2i}^G\ell_{2i}^2\right),\\
&		\gamma_{i1}^G=\frac{\chi_{i1}^{G}}{\chi_{i2}^{G}}+\frac{\chi_{1i}^{G}}{\chi_{12}^{G}}, \; 
		\gamma_{12}^G=\frac{\chi_{12}^{G}}{\chi_{1i}^{G}}+\frac{\chi_{21}^{G}}{\chi_{2i}^{G}}, \; 
		\gamma_{2i}^G=\frac{\chi_{2i}^{G}}{\chi_{21}^{G}}+\frac{\chi_{i2}^{G}}{\chi_{i1}^{G}}. \; 
	\end{split}
\end{eqnarray}
The sum over triangles $\sum_{\triangle}$ is replaced by the sum over bonds $\sum_{ij}$ due to the fact that bond $ij$ is shared with the two triangles:
\begin{eqnarray}
	\label{discrete-H1-bond-triangle}
	\begin{split}
	&H_1=
		\sum_{ij} \left(\gamma_{ij}^G +\gamma_{ji}^G\right)\ell_{ij}^2,\\
		&\gamma_{ij}^G=\frac{\chi_{ij}^{G}}{\chi_{i1}^{G}}
+\frac{\chi_{ij}^{G}}{\chi_{i2}^{G}}, \quad \gamma_{ji}^G=\frac{\chi_{ji}^{G}}{\chi_{j1}^{G}}+\frac{\chi_{ji}^{G}}{\chi_{j2}^{G}}, \quad (\triangle),
	\end{split}
\end{eqnarray}
where bond $ij$ is shared with triangles $ij1$ and $ji2$.

\subsection{Discrete Hamiltonian on 3D cubic lattice \label{App-A-cubic}}
As shown in the preceding subsections in this Appendix, the difference in the lattice structure between square and triangular lattices is reflected in $\gamma_{ij}^G$. In contrast, in 3D case, it should be emphasized that the expression of $\gamma_{ij}^G$  as well as the lattice structure differs from the 2D cases. Therefore, in this section, we show how the 3D discrete expression of $H_1$ in Eq. (\ref{discrete-Hamiltonian})  is obtained from the continuous Hamiltonian
\begin{eqnarray}\label{continuous-H1}
	H_1=\int \sqrt{g^G}d^3x g_G^{ij}\frac{\partial \vec{r}}{\partial x^i}\cdot\frac{\partial \vec{r}}{\partial x^j}, 
\end{eqnarray}
where $g_G^{ij}$ is the Finsler metric given by (Fig. \ref{fig-A-1}(d))
\begin{eqnarray}
	\label{F-metric-3D}
	\begin{split}
		&g_{ij}^G=\begin{pmatrix}
			(\chi_{i1}^G)^{-2} & 0 & 0\\
			0 & (\chi_{i2}^G)^{-2} & 0 \\
			0 & 0 & (\chi_{i3}^G)^{-2}
		\end{pmatrix},\\
		&\sqrt{g^G}=\sqrt{\det g_{ij}^G}=\frac{1
		}{\chi_{i1}^{G}\chi_{i2}^{G}\chi_{i3}^{G}}, \quad g^{ij}_G=(g_{ij}^G)^{-1}.
	\end{split}
\end{eqnarray}
The $\chi_{ij}^G, (j\!=\!1,2,3)$, defined by Eq. (\ref{chi-G}), 
represent the unit Finsler length from vertex $i$ along the three different axes $x^j, (j\!=\!1,2,3)$. 

Replacing the integral with the sum over cubes and the differential with the difference according to Eq. (\ref{discretization}), we obtain  $\sum_\square \sqrt{g^G} \left(g_G^{i1}\ell_{i1}^2\!+\!g_G^{i2}\ell_{i2}^2\!+\!g_G^{i3}\ell_{i3}^2\right)$, where the local coordinate $(x^1,x^2,x^3)$ is assumed as shown in Fig. \ref{fig-A-1}(d). Using the expressions in Eq. (\ref{F-metric-3D}), we have
\begin{eqnarray}
	\label{discrete-H1-org}
	\begin{split}
		H_1&\to\sum_\square \frac{1
		}{\chi_{i1}^{G}\chi_{i2}^{G}\chi_{i3}^{G}} \left((\chi_{i1}^G)^{2}\ell_{i1}^2+(\chi_{i2}^G)^2\ell_{i2}^2+(\chi_{i3}^G)^2\ell_{i3}^2\right)\\
		&=\sum_\square  \left(\frac{\chi_{i1}^{G}
		}{\chi_{i2}^{G}\chi_{i3}^{G}}\ell_{i1}^2+\frac{\chi_{i2}^{G}
		}{\chi_{i3}^{G}\chi_{i1}^{G}}\ell_{i2}^2+\frac{\chi_{i3}^{G}
		}{\chi_{i1}^{G}\chi_{i2}^{G}}\ell_{i3}^2\right).
	\end{split}
\end{eqnarray}
There are eight possible local coordinates at vertex $i$ (Fig. \ref{fig-A-1}(d)), in contrast to four possible local coordinates in the case of two-dimensional square lattice as illustrated in Fig. \ref{fig-A-1}(b). 
The number "eight" is identical with the total number of vertices in a cube. This summation convention $\sum_\square$ for the local coordinates at vertex $i$ is equivalent with the sum over coordinates at  the eight vertices of cube.  
Therefore, summing over eight possible expressions of $H_1$ obtained by replacing $i123\!\to\!i243$, $i453$, $i513$, $i156$, $i546$, $i426$, $i216$ in Eq. (\ref{discrete-H1-org}), we have
$3\times 7$ additional terms such as
\begin{eqnarray}
	\label{discrete-H1-other-3D}
	\begin{split}
		&\frac{\chi_{i2}^{G}
		}{\chi_{i4}^{G}\chi_{i3}^{G}}\ell_{i2}^2+\frac{\chi_{i4}^{G}
		}{\chi_{i2}^{G}\chi_{i3}^{G}}\ell_{i4}^2+\frac{\chi_{i3}^{G}
		}{\chi_{i2}^{G}\chi_{i4}^{G}}\ell_{i3}^2\\
		+&\frac{\chi_{i4}^{G}
		}{\chi_{i5}^{G}\chi_{i3}^{G}}\ell_{i4}^2+\frac{\chi_{i5}^{G}
		}{\chi_{i3}^{G}\chi_{i4}^{G}}\ell_{i5}^2+\frac{\chi_{i3}^{G}
		}{\chi_{i5}^{G}\chi_{i4}^{G}}\ell_{i3}^2 \\
		+&\cdots\\
		+&\frac{\chi_{i2}^{G}
		}{\chi_{i1}^{G}\chi_{i6}^{G}}\ell_{i2}^2+\frac{\chi_{i1}^{G}
		}{\chi_{i2}^{G}\chi_{i6}^{G}}\ell_{i1}^2+\frac{\chi_{i6}^{G}
		}{\chi_{i1}^{G}\chi_{i2}^{G}}\ell_{i6}^2.
	\end{split}
\end{eqnarray}
Since the sum over cubes $\sum_\square$ is replaced by the sum over vertices  $\sum_i$  as mentioned above, we have  
\begin{eqnarray}
	\label{discrete-H1-tetra-3D}
	\begin{split}
		H_1=&
		\sum_i  \left(\gamma_{i1}^G\ell_{i1}^2+\gamma_{i2}^G\ell_{i2}^2+\gamma_{i3}^G\ell_{i3}^2+\gamma_{i4}^G\ell_{i4}^2+\gamma_{i5}^G\ell_{i5}^2+\gamma_{i6}^G\ell_{i6}^2\right)\\
		=&\sum_i\sum_{j(i)}\gamma_{ij}^G\ell_{ij}^2=\frac{1}{2}\sum_i\sum_{j(i)}\gamma_{ij}^G\ell_{ij}^2+\frac{1}{2}\sum_j\sum_{i(j)}\gamma_{jij}^G\ell_{ji}^2\\
		=&\frac{1}{2}\sum_i\sum_{j(i)}\left(\gamma_{ij}^G+\gamma_{ji}^G\right)\ell_{ij}^2
		=\sum_{ij}\left(\gamma_{ij}^G+\gamma_{ji}^G\right)\ell_{ij}^2\\
	=&\sum_{ij}\Gamma_{ij}^G\ell_{ij}^2,\quad \Gamma_{ij}^G=\gamma_{ij}^G+\gamma_{ji}^G,
	\end{split}
\end{eqnarray}
where $\gamma_{ij}^G\!\not=\!\gamma_{ji}^G$ in the final expression, and $\gamma_{ij}^G, (j=1,\cdots,6)$ are given by
\begin{eqnarray}
	\label{discrete-H1-coeff-3D}
	\begin{split}
		&\gamma_{i1}^G=\frac{\chi_{i1}^{G}}{\chi_{i2}^{G}\chi_{i3}^{G}}+\frac{\chi_{i1}^{G}}{\chi_{i3}^{G}\chi_{i5}^{G}}+\frac{\chi_{i1}^{G}}{\chi_{i5}^{G}\chi_{i6}^{G}}+\frac{\chi_{i1}^{G}}{\chi_{i6}^{G}\chi_{i2}^{G}}, \; \gamma_{i2}^G=\frac{\chi_{i2}^{G}}{\chi_{i3}^{G}\chi_{i1}^{G}}+\frac{\chi_{i2}^{G}}{\chi_{i1}^{G}\chi_{i6}^{G}}+\frac{\chi_{i2}^{G}}{\chi_{i6}^{G}\chi_{i4}^{G}}+\frac{\chi_{i2}^{G}}{\chi_{i4}^{G}\chi_{i3}^{G}}, \\
		&\gamma_{i3}^G=\frac{\chi_{i3}^{G}}{\chi_{i1}^{G}\chi_{i2}^{G}}+\frac{\chi_{i3}^{G}}{\chi_{i2}^{G}\chi_{i4}^{G}}+\frac{\chi_{i3}^{G}}{\chi_{i4}^{G}\chi_{i5}^{G}}+\frac{\chi_{i3}^{G}}{\chi_{i5}^{G}\chi_{i1}^{G}}, \; \gamma_{i4}^G=\frac{\chi_{i4}^{G}}{\chi_{i3}^{G}\chi_{i2}^{G}}+\frac{\chi_{i4}^{G}}{\chi_{i2}^{G}\chi_{i6}^{G}}+\frac{\chi_{i4}^{G}}{\chi_{i6}^{G}\chi_{i5}^{G}}+\frac{\chi_{i4}^{G}}{\chi_{i5}^{G}\chi_{i3}^{G}}, \\
		&\gamma_{i5}^G=\frac{\chi_{i5}^{G}}{\chi_{i1}^{G}\chi_{i3}^{G}}+\frac{\chi_{i5}^{G}}{\chi_{i3}^{G}\chi_{i4}^{G}}+\frac{\chi_{i5}^{G}}{\chi_{i4}^{G}\chi_{i6}^{G}}+\frac{\chi_{i5}^{G}}{\chi_{i6}^{G}\chi_{i1}^{G}}, \; \gamma_{i6}^G=\frac{\chi_{i6}^{G}}{\chi_{i2}^{G}\chi_{i1}^{G}}+\frac{\chi_{i6}^{G}}{\chi_{i1}^{G}\chi_{i5}^{G}}+\frac{\chi_{i6}^{G}}{\chi_{i5}^{G}\chi_{i4}^{G}}+\frac{\chi_{i6}^{G}}{\chi_{i4}^{G}\chi_{i2}^{G}}. 
	\end{split}
\end{eqnarray}
Note that the four terms are identical in each of $\gamma_{ij}^g$, and we have $\gamma_{i1}^G=\!4\frac{\chi_{i1}^{G}}{\chi_{i2}^{G}\chi_{i3}^{G}}$ for example. Note also that  $\gamma_{i1}^G\!=\!\gamma_{i4}^G$, $\gamma_{i2}^G\!=\!\gamma_{i5}^G$ and $\gamma_{i3}^G\!=\!\gamma_{i6}^G$. 
The sum of six terms in the first of Eq. (\ref{discrete-H1-tetra-3D}) is written as the sum over vertices $j$ connected to $i$; $\sum_{j(i)}$, in the second line.  The final expression in Eq. (\ref{discrete-H1-tetra-3D}) is given by the sum over bonds $ij$.

We should note that the six terms in the first of Eq. (\ref{discrete-H1-tetra-3D}) correspond to the bonds connected to vertex $i$, and the four terms of $\gamma_{ij}^G$ in Eq. (\ref{discrete-H1-coeff-3D}) imply that bond $ij$ is shared by four cubes. Therefore, the number of terms in $\Gamma_{ij}^G\!=\!\gamma_{ij}^G\!+\!\gamma_{ji}^G$ in the sum $\sum_{ij}$ depends on whether vertices $i$, $j$ and bond $ij$ are inside or on the upper/lower surfaces. At vertex $i$ on the upper (lower) surface, there is no bond $i3$ ($i6$) (Fig. \ref{fig-A-1}(b)) for example. 
Since the sum over bonds $\sum_{ij}$ corresponds to the volume integral $\int \sqrt{g^G}d^3x$,  the terms corresponding to the surrounding cubes should be accounted for correctly  in the sum $\sum_{ij}$. At the upper (lower) surface bond $ij$, two cubes surround it, and correspondingly, $\gamma_{ij}^G$ in Eq. (\ref{discrete-H1-coeff-3D}) is given by two terms rather than four terms.

\subsection{Normalized effective tension coefficient of the Gaussian bond potential \label{App-A-2}}
As shown in the preceding subsection in the square lattice case, the summation convention can be changed from the sum over squares to the sum over bonds using the relation $\sum_\square \sum_{ij(\square)}\!=\!\sum_{ij}\sum_{\square(ij)}$, where $\sum_{\square(ij)}$ denotes the sum over squares sharing bond $ij$ (Fig. \ref{fig-A-1}(b)). Let $n_{ij}(=\!\sum_{\square(ij)}1)$ be the total number of squares sharing bond $ij$. Then, we have $n_{ij}\!=\!2$ for any bond $ij$, and $\sum_{\square(ij)}\gamma_{ij}^G$ is explicitly expressed in Eq. (\ref{discrete-H1-coeff}) for bond $ij$ and no symbol $\sum_{\square(ij)}$ is included in the final expression of $H_1$ in Eq. (\ref{discrete-H1-tetra}). 

The reason for introducing the normalization factor is to ensure that the expression $\sum_{ij}\Gamma_{ij}^G(\tau)\ell_{ij}^2$ is identical to the standard potential $\sum_{ij}\ell_{ij}^2$ for isotropic configurations $\tau^{\rm iso}$, which is at random everywhere. This condition $\Gamma_{ij}^G(\tau)\!\to\! 1 \;(\tau\!\to \!\tau^{\rm iso})$ is satisfied (Fig. \ref{fig-7}(b),(e)) in terms of the lattice average   $\frac{1}{\sum_{ij}1}\sum_{ij}\Gamma_{ij}^G(\tau)$, if $\Gamma_{ij}^G(\tau)$ is replaced by  $\frac{1}{\overline{\Gamma_{ij}^G(\tau^{\rm iso})}}\Gamma_{ij}^G(\tau)$, where $\overline{\Gamma_{ij}^G(\tau^{\rm iso})}$ denotes the mean value of $\Gamma_{ij}^G$ with $1\!\times\!10^5$ isotropic configurations of $\vec{\tau}$.   Therefore, we define the normalization factor $\bar{\Gamma}_G$ as follows:
\begin{eqnarray}
	\label{discrete-H1-bond-1}
\begin{split}
   &H_1=\sum_{ij} \Gamma_{ij}^G(\tau)\ell_{ij}^2, \\
   &\Gamma_{ij}^G(\tau)=\bar{\Gamma}_G^{-1} \left(\gamma_{ij}^G(\tau)+\gamma_{ji}^G(\tau)\right), \quad
   	\bar {\Gamma}_G=\frac{\sum_{ij}\overline{\gamma_{ij}^{G}(\tau^{\rm iso})+\gamma_{ji}^{G}(\tau^{\rm iso})}}{\sum_{ij}1}, \quad (\square).
\end{split}
\end{eqnarray}
The expression $\bar {\Gamma}_G$ in Eq. (\ref{discrete-H1-bond-1}) represents that $\bar {\Gamma}_G$ is the lattice average of  $ \overline{\gamma^{G}(\tau^{\rm iso})\!+\!\gamma_{ji}^{G}(\tau^{\rm iso})} $, which is defined by the mean value of $\gamma_{ij}^G\!+\!\gamma_{ji}^G$ using $1\!\times\!10^5$  isotropic configurations of $\vec{\tau}$ as described above. 

Due to the definition of $\bar {\Gamma}_G$ in Eq. (\ref{discrete-H1-bond-1}), we have $\sum_{ij}\Gamma_{ij}^G(\vec{\tau})/\sum_{ij}1\!\simeq\!1$ for any isotropic configuration $\vec{\tau}$. For the coefficient $D_{ij}^{u,v}$ in the Hamiltonians $H_{u,v}^D$ in Eq. (\ref{discrete-Hu-Hv}), the same expressions as that for $\Gamma_{ij}^G$ is used:  
\begin{eqnarray}
	\label{discrete-HuvD}
	\begin{split}
		&H_u^D=\sum_{ij}D_{ij}^u(\tau)\left(u_i-u_j\right)^2, \\
		&D_{ij}^u(\tau)={{\bar \Gamma}_u^{-1}}\left( \gamma_{ij}^u(\tau)+\gamma_{ji}^u(\tau)\right),   \quad 
		\bar {\Gamma}_u=\frac{\sum_{ij} \overline{ \gamma_{ij}^{u}(\tau^{\rm iso})+\gamma_{ji}^{u}(\tau^{\rm iso})}}{\sum_{ij}1},\\ 
		&H_v^D=\sum_{ij}D_{ij}^v(\tau)\left(v_i-v_j\right)^2, \\
		&D_{ij}^v(\tau)={{\bar \Gamma}_v^{-1}}\left( \gamma_{ij}^v(\tau)+\gamma_{ji}^v(\tau)\right),    	\quad
		\bar {\Gamma}_v=\frac{\sum_{ij} \overline{ \gamma_{ij}^{v}(\tau^{\rm iso})+\gamma_{ji}^{v}(\tau^{\rm iso})}}{\sum_{ij}1},
	\end{split}
\end{eqnarray}

\subsection{Discrete Laplacian \label{App-A-3}}
We describe the discrete expression of Laplacian $\Laplace u$  in this subsection. From the discrete expression of $H_u^D$,  $\Laplace u$ is obtained by the variation of  $H_u^D\!=\!\sum_{ij} D_{ij}^u(u_i-u_j)^2$ in Eq. (\ref{discrete-Hu-Hv}). Using the relation $\sum_{jk}\!=\!\frac{1}{2}\sum_j\sum_{k(j)}$, which describes the replacement of the sum over bonds $\sum_{jk}(=\!\sum_{kj})$ with the sum over vertices $\sum_j$ and the sum over the vertices $\sum _{k(j)}$ linked to the vertex $j$, we obtain
\begin{eqnarray} 
	\label{variation-discrete}
	\begin{split}
	\delta H_u^D=&\sum_i\delta u_i \frac{\delta}{\delta u_i} H_u^D=\sum_i\delta u_i \frac{\delta}{\delta u_i}\sum_{jk} D_{jk}^u(u_k-u_j)^2 \\
	=&2\sum_i\delta u_i \sum_{jk}D_{jk}^u\left(u_k\delta_{ki}-u_j\delta_{ki}-u_k\delta_{ji}+u_j\delta_{ji}\right) \\
	=&2\sum_i\delta u_i \sum_{jk}D_{jk}^u\left(u_k-u_j\right)\delta_{ki}+2\sum_i\delta u_i \sum_{kj}D_{kj}^u\left(-u_j+u_k\right)\delta_{ki} \\
	=&\sum_i\delta  u_i\sum_k\sum_{j(k)}D_{jk}^u\left(u_k-u_{j}\right)\delta_{ki}+\sum_i\delta u_i\sum_{k}\sum_{j(k)}D_{kj}^u\left(-u_{j}+u_{k}\right)\delta_{ki}\\
	=&2\sum_i\delta u_i \sum_{j(i)} D_{ji}^u\left(u_i-u_j\right).
	\end{split}
\end{eqnarray}
In the continuous expression of $H_u^D$, we obtain
\begin{eqnarray} 
	\label{variation-discrete-additional}
	\begin{split}
   \delta H_u^D({\rm cont})=&\delta \int \sqrt{g}d^2x g^{ab}\frac{\partial u}{\partial x^a}\frac{\partial u}{\partial x^b}\\
    =& \int \sqrt{g}d^2x g^{ab} \frac{\partial \delta u}{\partial x^a}\frac{\partial u}{\partial x^b}+\int \sqrt{g}d^2x g^{ab} \frac{\partial u}{\partial x^a}\frac{\partial \delta u}{\partial x^b}\\
    =&-  \int d^2x\delta u \frac{\partial}{\partial x^a}\sqrt{g}g^{ab}\frac{\partial u}{\partial x^b}-  \int d^2x \delta u \frac{\partial}{\partial x^b}\sqrt{g}g^{ab}\frac{\partial u}{\partial x^a}\\
    =&- 2 \int d^2x \delta u  \frac{\partial}{\partial x^a}\sqrt{g}g^{ab}\frac{\partial u}{\partial x^b}\\
    =&-2\int\sqrt{g} d^2x \delta u  \frac{1}{\sqrt{g}} \frac{\partial}{\partial x^a}\sqrt{g}g^{ab}\frac{\partial u}{\partial x^b}\\
	=&-2 \int\sqrt{g} d^2x \delta u\Laplace u.
	\end{split}
\end{eqnarray}
Thus,
under the correspondence between the continuous and discrete expressions of $H_u^D$ given by
\begin{eqnarray} 
	\label{continuous-discrete-correspondence}
 H_u^D({\rm cont})=\int \sqrt{g}d^2x g^{ab}\frac{\partial u}{\partial x^a}\frac{\partial u}{\partial x^b} \quad\leftrightarrow\quad H_u^D=\sum_{ij} D_{ij}^u(u_i-u_j)^2, 
\end{eqnarray}
we find the expression of the discrete Laplacian
\begin{eqnarray}
	\label{discrete-Laplace-App}
	\Laplace u_i = \sum_{j(i)} D_{ij}^u\left(u_j -u_i\right)= \left(\sum_{j(i)} D_{ij}^u u_j -u_i\sum_{j(i)} D_{ij}^u\right).
\end{eqnarray}
We should note that the standard network Laplacian $L_{ij}^{\rm std}$ can be extended as follows:
\begin{eqnarray}
	\label{discretization-Laplace}
	\begin{split}
L_{ij}^{\rm std}=A_{ij}-k_i\delta_{ij} \quad \rightarrow \quad L_{ij}=D_{ij}^u-\left(\sum_{k(i)}D_{ki}^u\right)\delta_{ij}
 	\end{split}
\end{eqnarray}
where $A_{ij}\!=\!1$ for linked vertices $ij$ and $A_{ij}\!=\!0$ otherwise, and $k_i\!=\!\sum_{j(i)}1$.

\section{Hamiltonians corresponding to the diffusion and reaction terms \label{App-B}}
In this Appendix, we derive Hamiltonians corresponding to the RD equation in Eq. (\ref{FN-eq}). 
 Starting with the continuous Hamiltonian 
\begin{eqnarray}
	\label{Continuous-Hamil-for-TP}
	\begin{split}
		&H=H_u+AH_v,\\
		&H_u=D_uH_u^D+H_u^R, \quad H_v=D_vH_v^D+H_v^R\\
		&H_u^D=\int \sqrt{g}d^2x g^{ab}\frac{\partial u}{\partial x^a}\frac{\partial u}{\partial x^b},\quad H_u^R=-\int \sqrt{g}d^2x\left(u^2-\frac{u^4}{2}-Buv\right),\\
		&H_v^D=\int \sqrt{g}d^2x g^{ab}\frac{\partial v}{\partial x^a}\frac{\partial v}{\partial x^b},\quad H_v^R=-\gamma\int \sqrt{g}d^2x\left(Buv-\alpha v^2-2\beta v\right),
	\end{split}
\end{eqnarray}
we show that the non-zero parameters $A$ and $B$ are uniquely determined for the Turing equation such that
\begin{eqnarray}
	\label{parameters-ABC}
	A=-\frac{1}{\gamma}(<0),\quad B=1.
\end{eqnarray}
$H_{u,v}^R$ denote the reaction terms corresponding to $f$ and $g$ in Eq. (\ref{FN-eq}).  $H_u$ and $H_v$ are written as
\begin{eqnarray}
	\label{Continuous-Hu-Hv-Hamil-for-TP}
	\begin{split}
		&H_u=\int \sqrt{g}d^2x \left(D_ug^{ab}\frac{\partial u}{\partial x^a}\frac{\partial u}{\partial x^b}-\left[u^2-\frac{u^4}{2}-Buv\right]\right),\\
		&H_v=\int \sqrt{g}d^2x \left(D_vg^{ab}\frac{\partial v}{\partial x^a}\frac{\partial v}{\partial x^b}-\gamma\left[Buv-\alpha v^2-2\beta v\right]\right),
	\end{split}
\end{eqnarray}
 By the variational technique with respect to $u$ and $v$, we have
\begin{eqnarray}
	\label{variation-u-v}
	\begin{split}
		0&=\delta H=\frac{\delta H}{\delta u}\delta	u+\frac{\delta H}{\delta v}\delta v\\ 
		&=\left(\frac{\delta H_u}{\delta u}+A\frac{\delta H_v}{\delta u}\right)\delta u+\left(\frac{\delta H_u}{\delta v}+A\frac{\delta H_v}{\delta v}\right)\delta v \\
		&=2\int \sqrt{g}d^2x \left(D_ug^{ab}\frac{\partial \delta u}{\partial x^a}\frac{\partial u}{\partial x^b}-\left[u-u^3-\frac{1}{2}Bv\right]\delta u-\frac{1}{2}A\gamma B v\delta u\right) \\
		&+2A\int \sqrt{g}d^2x \left(\frac{1}{2}A^{-1}Bu\delta v+D_vg^{ab}\frac{\partial \delta v}{\partial x^a}\frac{\partial v}{\partial x^b}-\gamma\left[\frac{1}{2}Bu-\alpha v-\beta\right]\delta v\right)\\
		&=2\int \sqrt{g}d^2x \left(-D_u \frac{1}{\sqrt{g}}\frac{\partial}{\partial x^a}\left(\sqrt{g}g^{ab}\frac{\partial  u}{\partial x^b}\right)-\left(u-u^3-\frac{1}{2}\left[B-AB \gamma\right]v\right)\right)\delta u \\
        &+2A\int \sqrt{g}d^2x \left(-D_v \frac{1}{\sqrt{g}}\frac{\partial}{\partial x^a}\left(\sqrt{g}g^{ab}\frac{\partial  v}{\partial x^b}\right)-\gamma \left(\frac{1}{2}\left[B-\frac{B}{A\gamma}\right]u-\alpha v-\beta\right)\right)\delta v
	\end{split}
\end{eqnarray}
for arbitrary variations $\delta u$ and $\delta v$. Thus, we obtain
\begin{eqnarray}
	\label{variation-u-v-2}
	\begin{split}
		&-D_u \frac{1}{\sqrt{g}}\frac{\partial}{\partial x^a}\left(\sqrt{g}g^{ab}\frac{\partial  u}{\partial x^b}\right)-\left(u-u^3-\frac{1}{2}\left[B-AB \gamma\right]v\right)=0, \\
		&-D_v \frac{1}{\sqrt{g}}\frac{\partial}{\partial x^a}\left(\sqrt{g}g^{ab}\frac{\partial  v}{\partial x^b}\right)-\gamma \left(\frac{1}{2}\left[B-\frac{B}{A\gamma}\right]u-\alpha v-\beta\right)=0. \\
	\end{split}
\end{eqnarray}
By letting  $B\!-\!AB \gamma\!=\!2$ and $B\!-\!\frac{B}{A\gamma}\!=\!2$, we find $A\!=\!-1/\gamma$ and $B\!=\!1$, and therefore
\begin{eqnarray}
	\label{FN-eq-curved-surface}
	\begin{split}
		&D_u \frac{1}{\sqrt{g}}\frac{\partial}{\partial x^a}\left(\sqrt{g}g^{ab}\frac{\partial  u}{\partial x^b}\right)+\left(u-u^3-v\right)=0, \\
		&D_v \frac{1}{\sqrt{g}}\frac{\partial}{\partial x^a}\left(\sqrt{g}g^{ab}\frac{\partial  v}{\partial x^b}\right)+\gamma \left(u-\alpha v-\beta\right)=0, 
	\end{split}
\end{eqnarray}
which are identical with the steady-state RD equation in Eq. (\ref{FN-eq}) for $\beta\!=\!0$.

	The final expressions of the continuous $H_u$ and $H_v$ are given by
\begin{eqnarray}
	\label{Continuous-Hamil-for-TP-final}
	\begin{split}
		&H=H_u-\frac{1}{\gamma}H_v,\\
		&H_u=D_uH_u^D+H_u^R, \quad H_v=D_vH_v^D+H_v^R\\
		&H_u^D=\int \sqrt{g}d^2x g^{ab}\frac{\partial u}{\partial x^a}\frac{\partial u}{\partial x^b},\quad H_u^R=-\int \sqrt{g}d^2x\left(u^2-\frac{u^4}{2}-uv\right),\\
		&H_v^D=\int \sqrt{g}d^2xg^{ab}\frac{\partial v}{\partial x^a}\frac{\partial v}{\partial x^b},\quad H_v^R=-\gamma\int \sqrt{g}d^2x\left(uv-\alpha v^2-2\beta v\right).
	\end{split}
\end{eqnarray}
The discrete expressions of $H_u$ and $H_v$ are given by Eq. (\ref{discrete-Hu-Hv}).

\section{Direction-dependent interaction coefficients and Hamiltonian \label{App-C}}
To define the directional energy localization, we introduce direction-dependent effective surface tension  and Gaussian bond potential in this subsection.  The direction-dependent quantities for the diffusion energies $H_u$ and $H_v$ have the same structure as those of $H_1$, and hence, we only discuss the quantities for $H_1$.  

The direction-dependent coefficient is defined by
\begin{eqnarray}
	\label{gamma-decomposition}
		\Gamma^{G}_\mu=\frac{1}{\sum_{ij}|\vec{e}_{ij}\cdot\vec{e}^{\,\mu}|}\sum_{ij}\Gamma_{ij}^G|\vec{e}_{ij}\cdot\vec{e}^{\,\mu}|,\quad (\mu=x,y).
\end{eqnarray}
The direction-dependent Gaussian bond potential is defined by
\begin{eqnarray}
	\label{H1-components}
		H_1^\mu=\frac{\sum_{ij}\Gamma_{ij}^G \ell_{ij}^2(\vec{e}_{ij}\cdot\vec{e}^{\,\mu})^2}{\Gamma_\mu^G}, \quad (\mu=x,y).
\end{eqnarray}
Using the relation $1\!=\!(\vec{e}_{ij}\cdot\vec{e}^{\,x})^2\!+\!(\vec{e}_{ij}\cdot\vec{e}^{\,y})^2$, it is easy to check that
\begin{eqnarray}
	\label{H1-decomposition}
	\begin{split}
		H_1=&\sum_{ij}\Gamma_{ij}^G\ell_{ij}^2=\sum_{ij}\Gamma_{ij}^G\ell_{ij}^2(\vec{e}_{ij}\cdot\vec{e}^{\,x})^2+\sum_{ij}\Gamma_{ij}^G\ell_{ij}^2(\vec{e}_{ij}\cdot\vec{e}^{\,y})^2\\
		=&\Gamma_x^G\frac{\sum_{ij}\Gamma_{ij}^G \ell_{ij}^2(\vec{e}_{ij}\cdot\vec{e}^{\,x})^2}{\Gamma_x^G}+\Gamma_y^G\frac{\sum_{ij}\Gamma_{ij}^G \ell_{ij}^2(\vec{e}_{ij}\cdot\vec{e}^{\,y})^2}{\Gamma_y^G}\\
		=&\Gamma_x^G H_1^x+\Gamma_y^G H_1^y,
	\end{split}
\end{eqnarray}
which is considered to be a macroscopic directional-decomposition of Gaussian bond potential. 

On the square lattice,
\begin{eqnarray}
	\label{gamma-decomposition-2}
		|\vec{e}_{ij}\cdot\vec{e}^{\,\mu}|=\left\{ \begin{array}{@{\,}ll}
			1\quad  (\vec{e}_{ij}\parallel\vec{e}^{\,\mu}) \\
			0\quad  (\vec{e}_{ij}\perp \vec{e}^{\,\mu})
		\end{array} \right.,\quad (\mu=x,y),\quad({\rm \square})
\end{eqnarray}
is satisfied because the bond direction $\vec{e}_{ij}$ is parallel or perpendicular to the canonical basis vectors $\vec{e}^\mu, (\mu=x,y)$. In this case, $\sum_{ij}|\vec{e}_{ij}\cdot\vec{e}^\mu|$ is the total number of bond along the $\mu$ direction, which is the half of the total number $N_B$ of bonds. For this reason, $\Gamma_\mu^G$ in Eq. (\ref{gamma-decomposition}) and $\Gamma_\mu^GH_1^\mu$ in Eq. (\ref{gamma-decomposition-2}) are respectively identical with the lattice average of the coefficient $\Gamma_{ij}^G$  and the energy $\Gamma_{ij}^G\ell_{ij}^2$ on bond $ij$ along the $\mu$-direction on the square lattice. For this reason, it is reasonable to assume $\Gamma_\mu^G$ and $\Gamma_\mu^GH_1^\mu$  as a direction-dependent coefficient and a directional component of energy, respectively, on the triangulated lattice. 

Small symbols are used to denote the energy per bond for the data plot, such as
\begin{eqnarray}
	\label{energy-per-bond}
	\begin{split}
&h_1={H_1}/{N_B},\\
&h_1^{\mu}=\frac{H_1^\mu}{\sum_{ij}(\vec{e}_{ij}\cdot\vec{e}^{\,\mu})^2}=\frac{1}{\sum_{ij}(\vec{e}_{ij}\cdot\vec{e}^{\,\mu})^2}\frac{\sum_{ij}\Gamma_{ij}^{G}\ell_{ij}^2(\vec{e}_{ij}\cdot\vec{e}^{\,\mu})^2}{\Gamma_\mu^G},\quad (\mu=x,y)
	\end{split}
\end{eqnarray}
for both 2D triangular and square lattice models and 3D cubic model.

	\section{On the lattice spacing  \label{App-D}}
We find from Figs. \ref{fig-7}(b),(e) that
	\begin{eqnarray}
		\label{h1-sq-tri}
		h_1=\frac{H_1}{N_B}\simeq\left\{ \begin{array}{@{\,}ll}
			0.515 \;(R\to 1)\quad  (\square)\;  \\
			0.336 \;(R\to 1)\quad  (\triangle)\;  
		\end{array} \right.
		\Leftrightarrow H_1\simeq 
		N \;(R\to 1)\quad  (\square,\; \triangle) 
	\end{eqnarray}
where $N_B$ is the total number of bonds, and $N_B\!=\!2N$ ($N_B\!=\!3N$) for the square (triangular) lattice. From Fig. \ref{fig-9}(b), we find 	
\begin{eqnarray}
		\label{h1-cube}
		h_1=\frac{H_1}{N_B}\simeq 0.529\;(R\to 1) \Leftrightarrow H_1\simeq \frac{3N}{2}\;(R\to 1) \quad ({\rm cube}),
\end{eqnarray}
where $N_B=3(n_x \!\times\! n_y\!\times\! (n_z-1))\!+\!2(n_x\!\times\! n_y)\!=\!27,200$ for the cubic lattice in Fig. \ref{fig-2}(c) due to the free boundary condition along the $z$ axis. Therefore,  $N_B/N\!=\!27,200/9,600\!\simeq 2.833$, and we obtain $H_1/N\!\simeq\! 1.5$. 
Thus, we find from the numerical data that  
\begin{eqnarray}
		\label{h1-sq-tri-cube}
		{H_1}\simeq \frac{DN}{2}\; (R\to 1),\quad (D=2,3),
\end{eqnarray}
for the square and triangular (in ${\bf R}^D, D\!=\!2$) and cubic (in ${\bf R}^D, D\!=\!3$) lattices.  Note that the relation $H_1\!=\! \frac{DN}{2}$ always holds for fluctuating lattices without fixed boundaries in ${\bf R}^D$, as a consequence of the scale invariance of the partition function $Z$ \cite{Wheater-JPA1994}. This property plays a crucial role in deriving the free energy and entropy of fixed lattices, as described in the Supplementary Material. 

Note that the value of $H_1$ depends on the lattice spacing $a$ on non-fluctuating lattices because the corresponding partition function does not possess scale invariance. As $a$ increases, then so does $H_1$, and vice versa. Consequently, the relative contribution of $H_1$ to the total Hamiltonian $H$, and hence to the IDOF $\vec{\tau}$, is expected to increase with increasing $a$ and decrease with decreasing $a$. For this reason, we chose the values of $a$ given in Eqs. (\ref{side-length-2D}) and (\ref{side-length-3D}) such that ${H_1}\!=\! \frac{DN}{2}$ is approximately satisfied.  Furthermore, the relation in Eq. (\ref{h1-sq-tri-cube}), which is determined solely  by the lattice spacing $a$  for  non-fluctuating lattices, is independent of $\chi_0$ owing  to the normalization factors $\bar{\Gamma}_G$ appearing in the intensive component $\Gamma_{ij}^G$ of $H_1$ in Eq. (\ref{discrete-H1-bond-1}).

We now discuss the implications of the relation observed in Eq. (\ref{h1-sq-tri-cube}). In contrast to non-fluctuating lattices, the value of $a$ on fluctuating lattices can be identified with the mean bond length $\ell$, which is automatically determined by the relation ${H_1}\!=\! \frac{DN}{2}$  for fluctuating lattices without fixed boundaries. By approximating $\ell_{ij}^2$ in $H_1\!=\!\sum_{ij}\ell_{ij}^2$ by the mean value ${\ell}^2$, we obtain $H_1\!=\!\sum_{ij}\ell^2\!=\!\ell^2\sum_{ij}1\!=\!N_B\ell^2\!=\!N_B\,a^2$ for the standard surface model, which corresponds to the FG surface model in the limit of $\chi_0\!\to\!\infty$. Therefore,  for $D\!=\!2$, the relation $H_1\!=\!\frac{DN}{2} \!=\!N_B a^2\!=\!3Na^2$ together with $N_B\!=\!3N$, yields $a\!=\!1/\sqrt{3}\!\simeq\!0.577$  on fluctuating triangular surface.  This value $a\!\simeq\!0.577$ is very close to the assumed one $a\!=\!0.58$ in Eq. (\ref{side-length-2D}). Thus,  the lattice spacing $a$ for non-fluctuating 2D surfaces and the 3D cube is chosen so that  the energy scale of $H_1\!=\!\sum_{ij}\Gamma_{ij}^G(\tau)\ell_{ij}^2$ is approximately the same as that of the fluctuating vertex model, for which the potential $H_1\!=\!\sum_{ij}\ell_{ij}^2$ is naturally defined. We emphasize that,  within the FG modeling framework, $H_1\!=\!\sum_{ij}\Gamma_{ij}^G(\tau)\ell_{ij}^2$ is well-defined on  both non-fluctuating and fluctuating lattices, in sharp contrast to the conventional potential $H_1\!=\!\sum_{ij}\ell_{ij}^2$, which is well-defined only for fluctuating lattices. It should also be noted that, in simulations of fluctuating vertex models without fixed boundaries, dimensionless curvature energies, such as the bending energy, are required to stabilize the surface shape, even though the mean bond length $\ell$ is essentially independent of whether the surface is in a smooth or crumpled phase \cite{KANTOR-NELSON-PRA1987,Gompper-Kroll-PRA1992,HELFRICH-1973,Polyakov-NPB1986,Peliti-Leibler-PRL1985,Bowick-PRep2001,NELSON-SMMS2004,Wheater-JPA1994,KOIB-PRE-2005}.


\begin{thebibliography}{9}
\bibitem{note-1}
	The variable $\vec{\tau}$, numerically generated on the unit circle or unit sphere, is  used  in the Hamiltonian only in the form of  $|\vec{\tau}|$  or  $(\vec{\tau})^2$, and therefore, $\vec{\tau}$ is effectively restricted to a  half circle or half sphere.
\end{thebibliography}

\end{document}